\documentclass[11pt]{amsart}
\usepackage[thin=0.5pt,dpi=1270]{diagrams}
\usepackage{geometry}
\usepackage{setspace}
\usepackage{hyperref}
\usepackage{lscape}
\newcommand\uu[2]{\ensuremath{U_{#1,#2}}}
\newcommand\Top{\rule{0pt}{2.6ex}}
\newcommand\Bot{\rule[-1.2ex]{0pt}{0pt}}
\newcommand\tb{\Top \Bot}
\newcommand{\del}[0]{\partial}
\newcommand{\abs}[1]{\vert #1\vert}
\newcommand{\takes}[2]{\!:\!#1 \rightarrow #2}
\newcommand{\sheaf}[1]{\ensuremath{\mathcal{#1}}}
\newcommand{\ra}[0]{\ensuremath{\rightarrow}}
\newcommand{\lra}[0]{\ensuremath{\longrightarrow}}
\newcommand{\nlra}[1]{\stackrel{#1}{\lra}}
\newcommand{\ZZ}[0]{\ensuremath{\mathbb{Z}}}
\newcommand{\CC}[0]{\ensuremath{\mathbb{C}}}
\newcommand{\PP}[0]{\ensuremath{\mathbb{P}}}
\newcommand\pone{{\ensuremath{\PP^1}}}
\newcommand\pones{{\ensuremath{\PP^1\times\PP^1}}}
\newcommand\wt{\widetilde}
\newcommand\pmn{\ensuremath{\PP^{2m+1} \times \PP^{2n+1}}}
\newcommand\dual{\ensuremath{*}}
\newcommand\defMatrix{
\ensuremath{ \left ( 
  \begin{matrix}
                x_0 & x_1 &\gamma_1 y_0 + \gamma_2 y_1 &\gamma_3 y_0 \\
                \epsilon_1 x_0 + \epsilon_2 x_1 & \epsilon_3 x_0 & y_0 & y_1 
         \end{matrix} 
         \right) }}

\begin{document}
\bibliographystyle{halpha}

\title{Deformed Quantum Cohomology and \protect \( (0,2) \protect \) Mirror Symmetry}

\author[J.~Guffin]{Josh Guffin\protect \({}^\dagger\protect \)}
\address{\protect \({}^\dagger\protect\)Department of Physics, 
1110 W. Green St., MC-704,
University of Illinois at Urbana-Champaign,
Urbana, IL 61801}
\email{guffin@uiuc.edu}

\author[S.~Katz]{Sheldon Katz\protect \({}^\ddagger\protect \)}
\address{\protect \({}^\ddagger\protect \)Department of Physics, 
1110 W. Green St., MC-704,
University of Illinois at Urbana-Champaign,
Urbana, IL 61801}
\address{Department of Mathematics,
1409 W. Green St., MC-382,
University of Illinois,
Urbana, IL  61801 }
\email{katz@math.uiuc.edu}

\date{October 2007}

\begin{abstract}
  We compute instanton corrections to correlators in the genus-zero
  topological subsector of a $(0,2)$ supersymmetric gauged linear sigma
  model with target space $\pones$, whose left-moving fermions  couple to a
  deformation of the tangent bundle.  We then deduce the theory's chiral
  ring from these correlators, which reduces in the limit of zero
  deformation to the $(2,2)$ ring.  Finally, we compare our results with
  the computations carried out by Adams et al.\cite{Adams:2003zy} and Katz
  and Sharpe \cite{Katz:2004nn}.  We find immediate agreement with the
  latter and an interesting puzzle in completely matching the chiral ring
  of the former.
\end{abstract}

\vspace*{1in}

\begin{flushright}
  \begin{tabular}{r}
         ILL-TH-07-3\\
         hep-th: 0710.2354
  \end{tabular}
\end{flushright}

\vspace*{1in}

\maketitle
\newpage
\doublespacing
\tableofcontents
\onehalfspacing
\newpage

\section{Introduction}

Twisted non-linear sigma models (NLSM) with $\mathcal N = (2,2)$
supersymmetry have a rich and fascinating structure.  One particularly
interesting facet is that a subset of their chiral operators form a ring,
which is a quantum-corrected version of the classical cohomology ring of
the target space \cite{Witten:1988xj,Lerche:1989uy,Witten:1989ig}.
Generically, NLSMs with $(2,2)$ supersymmetry may be deformed to $(0,2)$
theories, with a ring of ground-state operators generalizing the $(2,2)$
chiral ring \cite{Adams:2003zy,Adams:2005tc}.  These operators and their correlation
functions have been recently discussed in
\cite{Sharpe:2005fd,Witten:2005px,Tan:2006qt,Sharpe:2006qd}.\\

Mathematically, the $(2,2)$ chiral ring is described by quantum cohomology.
Although most $(0,2)$ chiral rings do not comprise deformations of $(2,2)$
rings, we study a specific example fulfilling this condition, and comment
on the general class to which it belongs.  In spite of the reduced
supersymmetry,  the subsector of operators under consideration remains
topological\cite{Adams:2005tc}. In particular, this ring is a Frobenius
algebra, a fact which we will exploit in our computations in \S
\ref{ssec:qcohomCalculation}.\\

A particularly interesting kind of $(0,2)$ model describes a theory whose
left-moving fermions couple to a holomorphic vector bundle on the target
space.  Such $(0,2)$ heterotic theories, once believed to have no
interpretation as string vacua due to instabilities in instanton
sectors\cite{Dine:1986zy,Dine:1987bq,Distler:1986wm,Distler:1987ee}, have
been shown to in fact be stable and flow to a conformal field
theory\cite{Silverstein:1995re,Berglund:1995yu,Basu:2003bq,Beasley:2003fx}.
These theories may be twisted by a non-anomalous $U(1)$ current, with
anomaly cancellation as in \cite{Distler:1996tj,Beasley:2003fx}.  \\

In the following sections, we compute the chiral ring of a $(0,2)$ NLSM
coupled to a deformation $\sheaf E$ of the tangent bundle of $\pones$ by
utilizing a gauged linear sigma model (GLSM) description
\cite{Morrison:1994fr,Distler:1993mk}.  In \S \ref{sec:background} we
review the field content and operators for trivial and non-trivial
instanton sectors.  The bundle to which the left-moving fermions couple is
derived and shown to arise as the cokernel in a short exact sequence of
sheaves on the target space.  Finally, the instanton moduli space and the
sheaf induced upon it by fermionic couplings are described.  As this sheaf
is unobstructed, the GLSM correlators will have a simple interpretation in
the geometric phase as an integral over the GLSM moduli space of zero
modes, as in \cite{Aspinwall:1991ce}.\\

Each step in the algorithm for computation of individual correlators is
described in \S \ref{sec:algorithm}; finding \v Cech representatives of
operators, computing their wedge product, and application of the
isomorphism from the top \v Cech cohomology to $\CC$.  This isomorphism is
simply the chiral ring's Frobenius form.  Additionally, we discuss the
applicability of the algorithm to more general toric varieties.\\

Once we have calculated correlators for instanton sectors of overall degree
$\leq 2$, we compute the chiral ring by deducing a quadratic relation between
them.  We perform this analysis in \S\ref{sec:cohomologyRelations}, and offer
further evidence of our derived relations from the Coulomb branch of the GLSM.
Finally in \S \ref{sec:comparison}, we offer a comparison between our results
and those of \cite{Katz:2004nn} and \cite{Adams:2003zy}.

\section{Background}
\label{sec:background}

\subsection{Classical Operators}
\label{sec:classicalOperators}

For convenience, we review a modicum of necessary information about the
particular half-twisted $(0,2)$ GLSM considered
here.   GLSMs with $(2,2)$ supersymmetry are described in \cite{Morrison:1994fr,Witten:1993yc},
those with $(0,2)$ in \cite{Distler:1993mk}, and 
half-twisted heterotic GLSMs in \cite{Silverstein:1995re}.\\

Our model includes four bosonic fields $\phi_1,\ \phi_2,\ \phi_3,\ \phi_4$,
charged under $U(1)\times U(1)$ with respective charges $(1,0),\ (1,0),\
(0,1),\ (0,1)$.  The space of vacua is given by the vanishing of the
D-terms
\begin{equation}
|\phi_1|^2+|\phi_2|^2-r_1=0,\qquad |\phi_3|^2+|\phi_4|^2-r_2=0
\label{eq:sympred}
\end{equation}  
with FI terms $r_1,r_2$, leading in the usual way
to $M=\pones$ in the geometric phase
where $r_1$ and $r_2$ are positive. \\

Before twisting, we have right-moving fermions $\psi_i$, superpartners
of the $\phi_i$, as well as four left-moving fermions $\rho_1,\
\rho_2,\ \rho_3,\ \rho_4$.  The $\rho_i$ have the same charges as the
corresponding $\phi_i$, but are not superpartners of these fields.  The
superfields $\Gamma_i$, for which $\rho_i$ are the respective lowest
components, are not chiral.  Their deviation from chirality is measured by
functions $E_i$ of the superfields $\Phi_i$, whose lowest components are
the $\phi_i$, and other chiral superfields $\Sigma,\wt
\Sigma$\cite{Witten:1993yc,Adams:2003zy}:
\begin{equation}
  \overline D_+ \Gamma_i = \sqrt{2} E_i (\Phi_j,\Sigma, \wt \Sigma).
  \label{eq:originOfE}
\end{equation}
If there is $(2,2)$ supersymmetry, then the $\rho_i$ are the superpartners of the $\phi_i$
under the additional supersymmetries.  \\

In the geometric phase, the $\psi_i$ fill out the tangent bundle of $M$
after twisting.  It will be useful to be explicit about this:
consider the tangent bundle of $M$, described as a cokernel in the
exact sequence\footnote{All commutative diagrams and some exact sequences
in this paper were typeset using Paul Taylor's {\it diagrams} package,
available at \url{http://www.cs.man.ac.uk/~pt/diagrams/}.}
\begin{equation}
\begin{diagram}
  0 & \rTo & 
  \sheaf O \oplus \sheaf O &
  \rTo^{ \left ( {\small
    \begin{matrix}
                x_0 & 0 \\
                x_1 & 0 \\
                0 & y_0 \\
                0 & y_1 \\
         \end{matrix}}
         \right)
  }
  &
  \sheaf O (1,0)^2 \oplus \sheaf O (0,1)^2 &\rTo& T_M &\rTo& 0.
\end{diagram}
\label{eq:ponesles}
\end{equation}

We will write $\{x_0,x_1,y_0,y_1\}$ for the local coordinates on the moduli
space induced by the fields $\{\phi_1,\phi_2,\phi_3,\phi_4\}$: more details
about this relationship will be described in \S
\ref{ssec:instantonBackground}.  Here, $\sheaf O (m,n)$ is the sheaf on $M$
defined as $\pi_1^*\sheaf O(m) \otimes \pi_2^* \sheaf O(n)$, with $\pi_1$
and $\pi_2$ the natural projection maps to each $\pone$,  and  $\sheaf
O(m,n)^2 \equiv \sheaf O(m,n) \oplus \sheaf O(m,n)$.\\

Before imposing gauge equivalence, the $\psi_i$ fill out the bundle $\sheaf O
(1,0)^2 \oplus \sheaf O (0,1)^2$ on $M$, and \eqref{eq:ponesles} says that
after gauge equivalence is taken into account, the $\psi_i$ actually fill out
the quotient bundle $T_M$.\\

The situation for the $\rho_i$ is similar.  Prior to imposing fermionic
gauge symmetries, the $\rho_i$ fill out the bundle $\sheaf O(1,0)^2 \oplus
\sheaf O(0,1)^2$ on $M$ as before.  Subsequent to imposition, however, the
fermionic and ordinary gauge symmetries do not necessarily coincide;  in
fact $\{\rho_1,\rho_2,\rho_3,\rho_4\}$ fill out the quotient bundle $\sheaf
E$ of $\sheaf O(1,0)^2\oplus\sheaf O(0,1)^2$.  In the $(2,2)$ situation we
have that $\sheaf E=T_M$, but in general $\sheaf E$ is a deformation of
$T_M$. \\

The first cohomology group valued in the sheaf of endomorphisms of $T_M$
describes the space of all first-order deformations.
Since $T_M\simeq\sheaf O(-2,0)\oplus\sheaf O(0,-2)$, we compute that 
\[\textit{End}\;(T_M)\simeq \sheaf O(-2,2)\oplus \sheaf O(0,0)^2\oplus \sheaf O(2,-2),\]
leading to
\begin{equation}
  \begin{split}
         H^1(M, \textit{End}\;({T_M}))= & 
         \left[H^1(\pone,\sheaf O(-2))\otimes H^0(\pone,\sheaf O(2))\right]\\
         &\oplus \left[H^0(\pone,\sheaf O(2))\otimes H^1(\pone,\sheaf O(-2))\right],
  \end{split}
\end{equation} 
a six-dimensional complex vector space. 
We therefore introduce six complex parameters
$\{\epsilon_1,\epsilon_2,\epsilon_3,\gamma_1,\gamma_2,\gamma_3\}$ as a
basis for the space of deformations and a matrix 
\begin{equation}
  F \equiv \defMatrix,
  \label{eq:generalDeformationMatrix}
\end{equation}

which encodes the most general deformation $\sheaf E$ of the tangent
bundle as the cokernel of ${}^t\! F$:
\begin{equation}
  0 \lra \sheaf O \oplus \sheaf O \nlra {{}^t\!F}  \sheaf O (1,0)^2 \oplus \sheaf O (0,1)^2 \lra \sheaf E \lra 0.
  \label{eq:ses}
\end{equation}
In other words, the data of $F$ describes the fermionic gauge symmetries,
and the fermions $\rho_i$ fill out the bundle $\sheaf E$ after imposition
of the symmetries.  Physically, the functions $E_i$ in \eqref{eq:originOfE}
encode the matrix \eqref{eq:generalDeformationMatrix}, and thus the
deformation as well.  Careful readers may note that $\sheaf E$ is not
necessarily a bundle; we will explore this detail further in \S
\ref{sec:degenerations}, but for this section and the next we will assume
that it is.\\

The topological sector of the twisted theory,  described mathematically by the
cohomology groups $H^p(\Lambda^q \sheaf E^*)$ \cite{Distler:1987ee}, is
comprised of operators corresponding to massless Ramond-Ramond states in the
untwisted theory.  In the twisted theory, massless states are expressed in
terms of the fields as 
\begin{equation}
  f_{J\Omega}(\phi)\psi^J\rho^\Omega,
  \label{eq:topOperators}
\end{equation}
where  $J$ and $\Omega$ are multi-indices for $T_M$ and $\sheaf E^*$,
respectively.  Correlation functions of products of these fields correspond
naturally to cup\slash wedge products of the cohomology representatives.
For our particular theory, the relevant operators are all elements of
$H^1(\Lambda^1 \sheaf E^*)$, and non-vanishing classical correlation
functions correspond to elements of $H^2(\Lambda^2 \sheaf E^*)$.\\

Note that the bundle $\sheaf E$, as a deformation of the tangent bundle,
satisfies 
\begin{equation}
  \Lambda^2 \sheaf E^* \cong K_M,
  \label{eq:sheafConstraint}
\end{equation}
so that $\Lambda^2 \sheaf E^* \cong \sheaf O(-2,-2)$ on $M$ 
and $H^2(\Lambda^2 \sheaf E^*) \cong H^2 (K_M) \cong \CC$.  Examination
of the exact sequence \eqref{eq:ses} further reveals that the bundle
satisfies the heterotic anomaly cancellation conditions $c_1(\sheaf E) =
c_1(T_M)$ and $\text{ch}_2(\sheaf E) = \text{ch}_2(T_M)$.\\

As we will be working with $\sheaf E^*$, the relevant sequences will be the
dual of \eqref{eq:ses},
\begin{equation}
  0
  \lra \sheaf E^*
  \lra \sheaf O (-1,0)^2 \oplus \sheaf O (0,-1)^2
  \nlra {F} \sheaf O \oplus \sheaf O
  \lra 0,
  \label{eq:dses}
\end{equation}
and its induced long exact sequence in cohomology, 
\begin{equation}
  \begin{split}
         0 \lra &H^0(\sheaf E^*)  \lra H^0 \big(\sheaf O (-1,0)^2 \oplus \sheaf O (0,-1)^2 \big) \lra
         H^0(\sheaf O \oplus \sheaf O) \\
           \lra &H^1(\sheaf E^*)  \lra H^1 \big(\sheaf O (-1,0)^2 \oplus \sheaf O (0,-1)^2 \big) \lra
                 H^1(\sheaf O \oplus \sheaf O ) \lra \cdots.
  \end{split}
  \label{eq:lesCohom}
\end{equation}

Since $\sheaf O (-1,0)^2 \oplus \sheaf O (0,-1)^2$ has neither global sections
nor degree 1 cohomology, we see that 
\begin{equation}
  H^1(\sheaf E^*) \cong H^0(\sheaf O\oplus \sheaf O) \cong \CC^2.
  \label{eq:liftIsomorphism}
\end{equation}
We explicitly construct this first isomorphism in \S \ref{sec:0instlift} while
finding \v Cech representatives of operators.

\subsection{Operators in Instanton Backgrounds}
\label{ssec:instantonBackground}

The preceding analysis discussed the zero-instanton (classical) case, when the
image of $\Sigma$ under the map $\varphi$ is homologous to a point in $M$.  We
would also like to compute correlation functions in the presence of non-trivial
instanton backgrounds.\\

A non-trivial instanton is a non-homologically trivial map.  The space of
algebraic maps $\pone\ra\pone$ is $\ZZ$ graded, with negative grading
corresponding to the empty set, and zero grading corresponding to the set of
trivial maps (that is, constant maps to a point).  For positive $n$, the set of
degree $n$ maps to $\pone$ consists of pairs $\{(\phi_1(z),\phi_2(z))\}$ of
homogeneous degree $n$ polynomials in the worldsheet variables collectively
denoted $z$.  Thus, for maps $\pone\ra\pones$ of bi-degree $(m,n)$,
$\phi_1$ and $\phi_2$ become sections of $\sheaf O_\pone (m)$, while
$\phi_3$ and $\phi_4$ become sections of $\sheaf O_\pone (n)$.\\

We write the moduli space of such maps as $\sheaf M$, and we can use the
$\phi_i$ to define local coordinates.  In terms of the worldsheet homogeneous
coordinates $z_0, z_1$, maps of instanton degree $m$ are written as 
\[
\phi_i(z_0,z_1) =  \sum_{j = 0}^m a_{ij} {z_0}^j {z_1}^{m-j}.
\]
Here, the $a_{ij}$ are complex numbers.  Imposing gauge equivalence, we see
that  in the geometric phase,  the pairs of polynomials $(\phi_1,\phi_2)$ and
$(\phi_3,\phi_4)$ are to be considered up to independent scalar
multiplications.  Thus, we combine the collections $a_{1,i}$ and $a_{2,j}$ as
the set $\{x_0,\cdots, x_{2m+1}\}$ and the collections $a_{3,i}$ and $a_{4,j}$
as the set $\{y_0,\cdots, y_{2n+1}\}$.  By gauge equivalence, each collection
is defined up to independent scalar multiplications, so that the $x_i$ and
$y_j$ behave like homogeneous coordinates on the product of two projective
spaces.  We thus conclude that the degree $(m,n)$ moduli space is 
\[
\sheaf M = \PP^{2m+1}\times\PP^{2n+1}.
\]
Following the customary notational abuse, we use non-linear sigma model
language and think of $((\phi_1,\phi_2),(\phi_3,\phi_4))$ as a map from $\pone$
to $\pone\times\pone$ of degree $(m,n)$, although the map does not exist at
points of the worldsheet where either $\phi_1$ and $\phi_2$ or $\phi_3$ and
$\phi_4$ have simultaneous zeros.\\

In an instanton background, the operators of the theory become sections of
sheaves on the instanton moduli space $\sheaf M$: expanding out $\rho_1$
and $\rho_2$ in terms of the instanton moduli space coordinates, we see
that they in fact become sections of a sheaf $\sheaf F$ on $\sheaf M$.
There is a natural description of $\sheaf F^*$ in terms of a short
exact sequence on $\sheaf M$,
\begin{equation}
  \begin{diagram}[scriptlabels]
         0 & \rTo & \sheaf F^* & \rTo & \sheaf O(-1,0)^{2(m+1)} \oplus
			\sheaf O(0,-1)^{2(n+1)}  & \rTo^{\;\;\;F_{mn}} & \sheaf O^2 & \rTo & 0,
  \end{diagram}
  \label{eq:modulises}
\end{equation}
where $F_{mn}$ is the $(2m+2n+4) \times 2$ matrix 
\begin{equation}
  \small
  \begin{split} &
         F_{mn} = \left (
         \begin{matrix}
                x_0 & x_1 & \cdots & x_{m} & x_{m+1} & \cdots & x_{2m+1}\\ \epsilon_1 x_0 + \epsilon_2 x_{m+1} & \epsilon_1
                x_1 + \epsilon_2 x_{m+2} & \cdots       & \epsilon_1 x_{m+1} + \epsilon_2 x_{2m+1}& \epsilon_3 x_0 & \cdots &
                \epsilon_3 x_{m}
         \end{matrix} \right . \\
          & \qquad\qquad\qquad\left. 
         \begin{matrix} 
                \gamma_1 y_0 + \gamma_2 y_{n+1} & \gamma_1 y_1 + \gamma_2 y_{n+2} & \cdots       & \gamma_1 y_{n+1} +
                \gamma_2 y_{2n+1}& \gamma_3 y_0 & \cdots & \gamma_3 y_{n}\\
                y_0 & y_1 & \cdots & y_{n} & y_{n+1} & \cdots & y_{2n+1}
         \end{matrix} 
         \right ),
  \end{split}
  \label{eq:defmat}
\end{equation}
where $x_i$ (resp. $y_j$) are coordinates on $\PP^m$ (resp. $\PP^n$).\\

Morally, in non-linear sigma model language, the sheaf $\sheaf F$ is defined
on the moduli space $\sheaf M = \pmn$ in terms of the evaluation map $\text{ev}
\takes {\Sigma \times \sheaf M} M$ and the projection $\pi \takes {\Sigma
\times \sheaf M} \sheaf M$ to be 
\[
\sheaf F \equiv \pi_* \text{ev}^* \sheaf E.
\]
Note that the matrices \eqref{eq:generalDeformationMatrix} and \eqref{eq:defmat} 
are of equal rank.  As alluded to earlier, we will see that for some
values of the parameters, $\sheaf E$ ceases to be locally free.  For such 
values, $\sheaf F$ also ceases to be a bundle.\\

The stalk of this sheaf at a  map $\varphi$ is $H^0(\Sigma, \varphi^*\sheaf
E)$.  As explained in \S 5.3 of \cite{Katz:2004nn}, this is the correct sheaf
to describe operators as sections on the moduli space.  One should take this
statement with a grain of salt, since as noted earlier, the evaluation map is
not defined everywhere.\\

When dealing with the unobstructed case, \(R^1 \pi_* \text{ev}^* \sheaf E = R^1
\pi_* \text{ev}^*  T_M = 0\), we will have that
\begin{equation}
  \det \sheaf F^* = \Lambda^\text{top} \sheaf F^* \cong K_{\sheaf M},
\label{eq:anomcan}
\end{equation}
so that non-trivial instantons also satisfy the anomaly cancellation condition.
Note that in the linear sigma model description, \eqref{eq:anomcan} can be
verified directly: taking determinants in \eqref{eq:modulises} yields
\[\Lambda^\text{top} \sheaf F^* \cong \sheaf O(-2m-2,-2n-2) \cong K_{\sheaf M}.\]

As with classical operators, correlation functions in instanton backgrounds
correspond to cup\slash wedge products of the cohomology representatives.
However, due to the larger $U(1)$ anomaly on the instanton moduli space, more
operator insertions are required to yield a non-vanishing correlator.  \\

\section{Elements of the Algorithm}
\label{sec:algorithm}

We now give an account of the various steps in the computation of
correlation functions.  As mentioned before, elements of $H^1(\sheaf E^*)$
comprise the space of topological operators of interest.  We construct \v
Cech representatives of these operators: in particular, we follow the
construction of the coboundary homomorphism (from the long exact sequence
in cohomology induced by \eqref{eq:dses}) to the penultimate step in order
to express operators as cochains in $C^1(\sheaf O(-1,0)^2 \oplus \sheaf O
(0,-1)^2)$.  By omitting the final step, finding an element of $H^1(\sheaf
E^*)$, we will obtain significant computational simplification when
computing correlation functions.\\

We also work out the injection from $C^\text{top}(\det \sheaf E^*)$ to
$C^\text{top}(\det (\sheaf O(-1,0)^2 \oplus \sheaf O (0,-1)^2))$, in the form
of factors that must be divided from the correlation functions.  Finally, we
construct the isomorphism $H^\text{top}(\det \sheaf E^*) \cong H^\text{top}(K)
\cong \CC$, which evaluates the correlator. \\

For the purposes of \v Cech cohomology, we will take the usual algebraic cover
of $\PP^n$ by open sets $U_i = \{ [(x_0,x_0,\cdots,x_n)] \in\PP^n \vert x_i \ne
0\}$, and cover $\PP^m\times \PP^n$ by open sets
\newcommand\uc[4]{\ensuremath{\uu #1 #2 \cap \uu #3 #4}}
\begin{equation}
  \uu ij \equiv \{([x_0,x_1,\cdots,x_m], [y_0,y_1,\cdots, y_n]) \in
  \PP^m\times\PP^n \vert x_i \ne 0 \text{ and }
  y_j \ne 0\}.
  \label{eq:pmnOpenCover}
\end{equation}

\subsection{Classical Lifts}
\label{sec:0instlift}

In this section, we carry out the steps in the construction of the coboundary
map \eqref{eq:liftIsomorphism}.  Consider the complex of short exact sequences
\begin{equation}
  \begin{diagram}
         0 & \rTo & C^0(\sheaf E^*) & \rTo^a & C^0 \big(\sheaf O (-1,0)^2 \oplus \sheaf O (0,-1)^2 \big) &
         \rTo^F & C^0(\sheaf O \oplus \sheaf O) & \rTo & 0\\
         && \dTo^{\delta_0} && \dTo^{\delta_0}&& \dTo^{\delta_0}\\
         0 & \rTo  & C^1(\sheaf E^*) & \rTo^a  & C^1 \big(\sheaf O (-1,0)^2 \oplus \sheaf O (0,-1)^2 \big) &
         \rTo^F  & C^1(\sheaf O \oplus \sheaf O ) &\rTo &0 . \\
         && \dTo^{\delta_1} && \dTo^{\delta_1}&& \dTo^{\delta_1}\\
         && \vdots && \vdots && \vdots 
  \end{diagram}
  \label{eq:dlescohom}
\end{equation}

Beginning with each of the elements $(0,1)$, $(1,0)\in \ker \delta_0 \subset
C^0(\sheaf O\oplus \sheaf O)$, we use the fact that the matrix $F$ defines a
surjective map to find an element of $C^0 \big(\sheaf O (-1,0)^2 \oplus \sheaf
O (0,-1)^2 \big)$ which lifts it.  That is, for each $U_i \subset M$, we must
find a column vector $V$ whose first (respectively, last) two rows are rational
functions of overall degree $-1$ in the first (respectively, second) $\pone$'s
variables; 
\begin{equation}
  \defMatrix \cdot V = \left ( \begin{matrix} 1 \\ 0 \end{matrix} \right),
         \label{eq:simultaneousEquations}
\end{equation}
and similarly for $(0,1)$.  See the end of this section for a further
discussion of this computation, and Appendix \ref{sec:lifts} for the exact form
of the lifts.\\

Given these elements of $C^0 \big(\sheaf O(-1,0)^2 \oplus \sheaf O(0,-1)^2
\big)$, we apply $\delta_0$ to obtain elements $Y$ and $\wt Y$ of $C^1
\big(\sheaf O(-1,0)^2 \oplus \sheaf O(0,-1)^2 \big)$.  By commutativity of the
diagram, and the fact that \( \delta_0 (1,0) = \delta_0 (0,1) = 0, \) we see
that these elements vanish upon application of $F$.  The chains $Y$ and $\wt Y$
are therefore in the image of $a$, so that there are elements $\psi$, $\wt \psi
\in C^1(\sheaf E^*)$ satisfying $a(\psi) = Y$, $a(\wt \psi) = \wt Y$. We will
not obtain the explicit form of $\psi$ and $\wt \psi$, opting rather to pursue
computations using the $C^1 \big( \sheaf O(-1,0)^2 \oplus \sheaf O(0,-1)^2
\big)$ representatives:  in \S \ref{sec:minormap} we will use the map $a$ to
find the proper element of $H^2(\Lambda^2 \sheaf E^*)$.\\

We now show that upon lifting to elements in $C^1(\sheaf E^*)$, we will in fact
obtain an element of $H^1(\sheaf E^*)$.  This is a standard diagram chase
argument, which we reproduce for convenience of the reader.  For the proof of
this statement, we add numerical indices to the maps $a$ and $F$, explicitly
indicating to which row in diagram \eqref{eq:dlescohom} we are referring.  Let
$\psi$ and $\wt \psi$ be elements of $C^1(\sheaf E^*)$ satisfying
\[a_1(\psi) = Y\qquad a_1(\wt \psi) = \wt Y.\]
Exactness of the diagram implies that $\delta_1 Y = \delta_1 \wt Y = 0$, while
commutativity tells us that $a_2 \circ \delta_1 (\psi) = a_2 \circ \delta_1
(\wt \psi) = 0$, so that by injectivity of $a$, $\delta_1 \psi = \delta_1 \wt
\psi$ = 0.  This tells us that $\psi, \wt \psi$ represent cohomology classes in
$H^1(\sheaf E^*)$.  We reiterate, our computational intent lies with the
elements $Y$ and $\wt Y$, not $\psi$ and $\wt \psi$.\\

Thus far, we have treated the lifts abstractly.  Let us mention a few details
of the analysis.  On each open set, we must construct a solution to the system
of simultaneous equations \eqref{eq:simultaneousEquations}.  Such a solution
will be a well-defined rational function in terms of local variables.  This
implies that only monomials in the non-vanishing variables on that open set may
appear in the denominator.\\

For the simple deformation that appeared in \cite{Katz:2004nn}, solutions were
constructable by hand.   In the fully deformed theory, computations are
considerably more complicated: a computer algebra system is called for.  We
attempted to use Mathematica\cite{Mathematica:52}, specifically its ``Solve''
function.  Unfortunately, Solve could not be instructed to restrict to rational
functions with monomials of certain variables in the denominator.  A custom
solver was written to account for this condition.

\subsection{Instanton Lifts}
\label{sec:instLifts}

In instanton backgrounds, the operators in the topological subsector
\eqref{eq:topOperators} become elements of $H^1(\sheaf M, \sheaf F^*)$ by
virtue of the reinterpretation of the $\rho$ fermions as sections of $\sheaf
F^*$.  Na\"ively, one might think that we must reapply the algorithm for
classical lifts to the new sheaf.  However, a trick allows us to use the degree
$(0,0)$ solutions as lifts in instanton sectors of arbitrary degree.  \\

We begin by constructing, as in \eqref{eq:dlescohom}, an exact sequence of
complexes from the data of the short exact sequence of sheaves
\eqref{eq:modulises}.  These complexes allow us find elements of
\[C^0\Big(\sheaf O(-1,0)^{2(m+1)}\oplus \sheaf O(0,-1)^{2(n+1)}\Big )\]
that lift basis elements of $C^0(\sheaf O^2)$ on each open set of $\pmn$.  The
deformation matrix $F_{mn}$ describing the sheaf $\sheaf F^*$ on $\pmn$ appears
in \eqref{eq:defmat}.\\

On the open set where $x_i \ne 0$ and $y_j \ne 0$, we can use the solution
from the degree $(0,0)$ case by noting that only columns $(i+1)$ and
$(m+2+i)$ 
columns of $F_{mn}$ involve $x_i$, and
only columns $[2m+2+(j+1 
)]$ and $[2m+2+(n+2+j 
)]$
involve $y_j$.  The other variables appearing those columns do not appear
elsewhere, so we obtain the desired lift by letting $\{x_0,x_1,y_0,y_1\}$
in the degree $(0,0)$ lift go to $\{x_{i+1}, x_{m+2+i}, y_{j+1}, y_{n+2+j}\}$ and
writing the new solution as a column vector 
\begin{equation*}
  \includegraphics{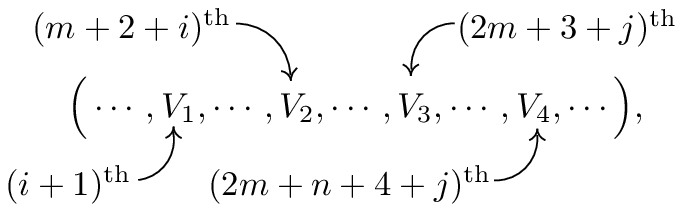}
\end{equation*}
where the elements of $V$ occupy the indicated positions, and all other
entries are zero.   With this realization, lifts are quickly and simply
computed on all open sets in the moduli space.

\subsection{The Minor Map}
\label{sec:minormap}

As noted before, the only non-vanishing correlators in the zero-instanton
sector are two-point functions.  The operator product inside the
correlation function must correspond to $\psi_i\cup \psi_j \in
H^2(\Lambda^2 \sheaf E^*)$, for some $\psi_i, \psi_j \in H^1(\sheaf E^*)$.
In higher instanton sectors, the condition for a correlator to be
non-vanishing will change.  Since we are examining a theory with fields
associated to an unobstructed sheaf, the appropriate number of operators
will simply be the complex dimension of the moduli space.\\

We first discuss the zero instanton case, before giving a general formula.
With our bundle restriction \eqref{eq:sheafConstraint}, we see that the second
exterior power $\Lambda^2 \sheaf E^* \cong \sheaf O(-2,-2)$.  Since we will be
dealing with representatives $Y$ and $\wt Y$ of $\psi$ and $\wt \psi$,
respectively, their cup\slash wedge  product will be an element of 
\[C^2\left (\Lambda^2 \left[\sheaf O(-1,0)^2 \oplus \sheaf O(0,-1)^2\right]\right) \cong C^2 (\sheaf O(-2,0) \oplus \sheaf O(-1,-1)^4 \oplus \sheaf O(0,-2)),\]
and we will obtain our final representative in $H^2\big(O(-2,-2)\big)$ by
specifying the inclusion
\[ i\!\!:{\sheaf O(-2,-2)} \hookrightarrow {\sheaf O(-2,0) \oplus \sheaf O(-1,-1)^4 \oplus \sheaf O(0,-2)}.\]
Any such map can be specified by giving an element of \(\sheaf O(0,2)
\oplus \sheaf O(1,1)^4 \oplus \sheaf O(2,0).\) How can we find the proper
element?  Since the diagram \eqref{eq:dses} is exact, the kernel of the
matrix defining the map between $\sheaf O(-1,0)^2 \oplus \sheaf O(0,-1)^2$
and $\sheaf O^2$
must contain the image of such an element.  This implies that 
the corresponding map in the exact sequence of
determinant bundles also contains the image, so that
on each open set the lifts will have a common factor --  an element
of \(\sheaf O(0,2) \oplus \sheaf O(1,1)^4 \oplus \sheaf O(2,0).\) \\

We compute this element, and thus the map, by finding the maximal minor
determinants of the kernel of \eqref{eq:defmat}.  A local basis
\(\{e_1,e_2,f_1,f_2\}\) for \(\sheaf O(0,1)^2 \oplus \sheaf O(1,0)^2\) induces
a basis \(\{e_1\wedge e_2, \cdots, f_1\wedge f_2\}\) for \(\sheaf O(0,2) \oplus
\sheaf O(1,1)^4 \oplus \sheaf O(2,0)\) on the same open set, so that the
explicit form of the map is given by the determinant of the appropriate maximal
minor multiplied by a basis element.  A basis for the kernel of
\eqref{eq:generalDeformationMatrix} is furnished by the column space of 
\begin{equation}
  \small
  \left(
  \begin{array}{ll}
         x_1 y_1-x_0 y_0 \gamma_3 \epsilon_3 & x_1 y_0-x_0 \left(y_0 \gamma_1+y_1 \gamma_2\right) \epsilon_3 \\
         -x_0 y_1+x_0 y_0 \gamma_3 \epsilon_1+x_1 y_0 \gamma_3 \epsilon_2
         & x_0 \left(y_1 \gamma_2 \epsilon_1+y_0 \left(\gamma_1 \epsilon_1-1\right)\right)+x_1 \left(y_0 \gamma_1+y_1 \gamma_2\right) \epsilon_2 \\
         0 & \epsilon_3 x_0^2-x_1 \epsilon_1 x_0-x_1^2 \epsilon_2 \\
         \epsilon_3 x_0^2-x_1 \epsilon_1 x_0-x_1^2 \epsilon_2 & 0
  \end{array}
  \right),
  \label{eq:deformationKernel}
\end{equation}
and the maximal minor determinants of this matrix are 
\begin{equation}
  \small
  \left\{
  \begin{array}{l}
         \left(\gamma _3 y_0^2-y_1 \gamma _1 y_0-y_1^2 \gamma _2\right)
         \left(\epsilon _3 x_0^2-x_1 \epsilon _1 x_0-x_1^2 \epsilon
         _2\right) ,\\
         -\left(\epsilon _3 x_0^2-x_1 \epsilon _1 x_0-x_1^2 \epsilon
         _2\right) \left(x_0 y_0 \gamma _3 \epsilon _3-x_1 y_1\right)
         ,\\
         \left(\epsilon _3 x_0^2-x_1 \epsilon _1 x_0-x_1^2 \epsilon
         _2\right)
         \left(x_0 \left(y_0 \gamma _1+y_1 \gamma _2\right)
         \epsilon _3-x_1
         y_0\right) ,\\
         -\left(x_0 \left(y_1-y_0 \gamma _3 \epsilon
         _1\right)-x_1 y_0 \gamma
         _3 \epsilon _2\right) \left(\epsilon _3 x_0^2-x_1
         \epsilon _1
         x_0-x_1^2 \epsilon _2\right) ,\\
         -\left(x_0 \left(y_1 \gamma _2 \epsilon
         _1+y_0 \left(\gamma _1
         \epsilon _1-1\right)\right)+x_1 \left(y_0
         \gamma _1+y_1 \gamma
         _2\right) \epsilon _2\right)
         \left(\epsilon _3 x_0^2-x_1 \epsilon
         _1 x_0-x_1^2 \epsilon _2\right) ,\\
         -\left(-\epsilon _3 x_0^2+x_1
         \epsilon _1 x_0+x_1^2 \epsilon
         _2\right)^2
  \end{array}
  \right \}.
\end{equation}
Note that each term contains a common factor \( -\epsilon_3 x_0^2+x_1 \epsilon_1 x_0+x_1^2 \epsilon_2 \).
Upon division by this polynomial, we find that the element of
\(\sheaf O(0,2) \oplus \sheaf O(1,1)^4 \oplus \sheaf O(2,0)\) encoding 
the desired map takes the form
\begin{equation}
  \begin{split}
         &\left(-\gamma_3 y_0^2+y_1 \gamma_1 y_0+y_1^2 \gamma_2\right) e_1\wedge e_2+
         \left(x_0 y_0 \gamma_3 \epsilon_3-x_1 y_1\right) e_1\wedge f_1\\
         &{}+ \left(x_1 y_0-x_0 \left(y_0 \gamma_1+y_1 \gamma_2\right) \epsilon_3\right) e_1\wedge f_2
         +\left[x_0 \left(y_1-y_0 \gamma_3 \epsilon_1\right)-x_1 y_0 \gamma_3 \epsilon_2\right] e_2\wedge f_1\\
         &{}+\left[x_0 \left(y_1 \gamma_2 \epsilon_1+y_0 \left[\gamma_1 \epsilon_1-1\right]\right)+x_1 \epsilon_2\left(y_0 \gamma_1+y_1 \gamma_2\right) \right] e_2\wedge f_2\\
         &{}+\left(\epsilon_3 x_0^2-x_1 \epsilon_1 x_0-x_1^2
         \epsilon_2\right) f_1\wedge f_2.
  \end{split}
  \label{eq:minormap}
\end{equation}
Each of the correlators we compute contains this term as a multiplicative
factor.\\

In order to actually compute a correlation function, we must take the cup\slash
wedge product of two cohomology representatives.  Since any such element will
be proportional to \eqref{eq:minormap}, we need only compute the $e_1 \wedge
e_2$ term.  This may be accomplished by first multiplying one of the
cohomological representatives by $f_1 \wedge f_2$, and then dividing by
$(-\gamma_3y_0^2 + y_1 \gamma_1 y_0 + y_1^2 \gamma_2)$.  Truncating the product
in this way relieves us of the computational burden of the other five wedge
coefficients, a savings that vastly increases in higher instanton sectors.   To
check, a few correlators were completely computed, and as expected were exactly
proportional to the map.\\

For higher-degree instanton sectors, the process applies \emph{mutatis
mutandis} to the kernel and minors of the matrix \eqref{eq:defmat}.

\subsection{Bundle Degenerations}
\label{sec:degenerations}

Let us emphasize that the parameter space spanned by the $\epsilon$'s and
$\gamma$'s describes deformations of the tangent \emph{sheaf}: there may be
configurations where $\sheaf E^\dual$ ceases to be a bundle.  Such a
configuration occurs precisely at the points on the instanton moduli space
where each of the minors vanishes identically at some point $p\in M$. At such a
point, the deformation matrix \eqref{eq:defmat} would no longer be of rank two:
the dimension of the stalk of $\sheaf E^\dual$ changes, so it cannot be a
bundle.\\

A polynomial in the deformation parameters detects the possibility of such
points.  We will restrict our attention to the open set in
$H^1(\mathrm{End}(T\sheaf M))$ where this polynomial does not vanish, so that
we have an honest bundle.\\

In order to detect the of points where the six deformation matrix minors
\begin{align}
  \nonumber
     &\epsilon_3 x_0^2-\epsilon_1 x_0x_1 -\epsilon_2x_1^2 & &x_0 y_0-\left(y_0 \gamma_1+y_1 \gamma_2\right) \left(x_0 \epsilon_1+x_1 \epsilon_2\right)\\
  \label{eq:minors}
     &x_0 y_1-y_0 \gamma_3 \left(x_0 \epsilon_1+x_1 \epsilon_2\right) & &x_1 y_0-x_0 \left(y_0 \gamma_1+y_1 \gamma_2\right) \epsilon_3\\
  \nonumber
     &x_1 y_1-x_0 y_0 \gamma_3 \epsilon_3  & &y_1 \left(y_0 \gamma_1+y_1 \gamma_2\right)-y_0^2 \gamma_3
\end{align}
vanish, we construct a Gr\"obner basis for the ideal $I$ generated by the
polynomials \eqref{eq:minors}, using an elimination order to eliminate the
variables $\{x_0,x_1,y_0,y_1\}$.  In other words, we choose an ordering of the
monomials so that the part of the Gr\"obner basis for $I$ which does not
include any of $\{x_0,x_1,y_0,y_1\}$ forms a basis for the intersection of $I$
with the subring of polynomials in the $\epsilon$'s and the $\gamma$'s.  The
resulting equations in the parameter variables define by construction the locus
of parameter values for which $\sheaf E$ fails to be a bundle at {\em some\/}
point of $\pones$, i.e.\ values for which $\sheaf E$ is not a bundle.\\

The computation is actually slightly more complicated.  The homogeneity of the
ideal $I$ requires extraneous factors of the variables to be eliminated in
order to get the degree high enough to achieve the desired elimination.  To
explain the result of the computation, we define a polynomial in the
deformation parameters, $\phi$:
\begin{equation}
  \phi \equiv \left(\gamma_1+\gamma_2 \gamma_3 \epsilon_1\right) \left(\epsilon_1+\gamma_1 \epsilon_2
     \epsilon_3\right)-\left(\gamma_2 \gamma_3 \epsilon_2 \epsilon_3-1\right)^2
  \label{eq:phiDef}
\end{equation}

Then we find that the only polynomial in $I$ consisting of powers of $x_1$ and
$y_1$ multiplied by a polynomial in the $\epsilon$'s and $\gamma$'s is
$x_1y_1\phi$.  Furthermore, we can compute directly that any common zeros of
the six minors satisfying either $x_1=0$ or $y_1=0$ necessarily satisfy
$\phi=0$.  The conclusion is that the locus in the moduli space parametrizing
sheaves $\sheaf E$ which are not bundles is precisely given by $\phi=0$.\\

We will find powers of $\phi$ in the denominators of some of our expressions
later in this paper.  This is not a concern, since we have just shown that
$\phi$ is nonzero whenever the theory is well-defined, so that $\sheaf E$ is
actually a bundle.

\subsection{\v Cech Cohomology of the Canonical Bundle}

Once we have interpreted our correlators as elements of $H^{\rm top} (\sheaf
M,\Lambda^{\rm top}\sheaf F^*)=H^{\rm top}(\sheaf M,K_{\sheaf M})$, the last
step is to evaluation the result as a number via a trace isomorphism
\begin{equation}
\mathrm{Tr}:H^{\rm top}(\sheaf M,K_{\sheaf M})\to \CC.
\end{equation}
We will describe how to find a trace map, passing to a broader context in which
$\sheaf M$ is replaced by an arbitrary smooth projective toric variety $X$.\\

We begin by recalling the description of homogeneous coordinates on $X$
\cite{cox:1995}.  Thinking of $X$ as being described by a fan $\Sigma$, we
associate variables $x_0,\ldots,x_N$ to the $N+1$ edges of
$\Sigma$.\footnote{If $X$ is the space of vacua of a GLSM, the set $\Sigma(1)$
of one-dimensional edges of $\Sigma$ is in one to one correspondence with the
set of chiral fields in the model.}  The homogeneous coordinate ring is the
polynomial ring generated by the $x_i$, and is
graded by the Chow group $A_{n-1}(X)$ of divisor classes on $X$, where $n= \dim
(X)$.  In this grading, the degree of $x_i$ is the class of the divisor $D_i$
defined by $x_i=0$.\\

Illustrating with projective space $\PP^n$, there are $n+1$ edges in the fan
and hence $n+1$ coordinates $x_0,\ldots,x_n$.  Each $D_i$ is a hyperplane,
of degree 1; so we assign each $x_i$ the degree~1 and we recover the usual
homogeneous coordinates on projective space. \\

The canonical class of $X$ is given by $K_X = - \sum_{i=0}^N D_i$, which we
denote as $-D$.  This description makes clear that $1/(x_0\cdots x_N)$ is a
(meromorphic) section of $K_X$.  The complement of $D$ is the torus
$T=(\CC^*)^n$, defined by $x_i\ne0$ for all $i$.\\

We know that $H^n(X,\sheaf O(-D))=H^n(X,K_X)$ is one dimensional, and we try to find a
generator.  We take the standard open cover of $X$ described by the top
dimensional cones in the fan.  Explicitly, if $\sigma$ is a top dimensional
cone, then the open set $U_\sigma$ is given as the locus where $x_i\ne0$, for
all $x_i$ corresponding to edges of the fan  {\em not\/} contained in $\sigma$.
For example, in the case $\PP^{2m+1}\times \PP^{2n+1}$ considered earlier, the
$U_\sigma$ are precisely the $U_{ij}$ defined in equation \eqref{eq:pmnOpenCover}.\\

Computing $H^n(X,\sheaf O(-D))$ by \v Cech cohomology using the above cover, we see that
each of the intersections $U_{\sigma_0}\cap\ldots\cap U_{\sigma_k}$ needed in
the \v Cech description necessarily contain the torus $T$.  Since the only
possible denominators of rational functions on $X$ which are holomorphic on all
of $T$ are monomials, we arrive at an important observation:\\

\smallskip
{\em In the \v Cech description of $H^n(X,\sheaf O(-D))$, we only need to consider
expressions which are sums of Laurent monomials.}\\
\smallskip

A {\em Laurent monomial\/} is a monomial where negative exponents are 
allowed.\\

A bit more pedagogically, we fix a multidegree $\mathbf{r}= (r_0,\ldots,r_N)$
and let $C^i(X,\sheaf O(-D))_\mathbf{r}$ denote those \v Cech $i$-cochains consisting
entirely of components which are scalar multiples of $x_0^{r_0}\cdots
x_N^{r_N}$.  Then we have\\
\begin{equation}
C^i(X,\sheaf O(-D))=\oplus_\mathbf{r}C^i(X,\sheaf O(-D))_{\mathbf{r}}.
\end{equation}

Since the coboundary maps
\[
\delta_i:C^i(X,\sheaf O(-D))\to C^{i+1}(X,\sheaf O(-D))
\]
preserve the degree of monomials, we get an induced decomposition
\begin{equation}
H^i(X,\sheaf O(-D))=\oplus_\mathbf{r}H^i(X,\sheaf O(-D))_{\mathbf{r}}.
\end{equation}

Note that since the monomials occurring in representatives of $H^n(X,\sheaf O(-D))$ must
be sections of $\sheaf O(-D)$, we see by considering the grading of the
homogeneous coordinate ring of $X$ that $H^n(X,\sheaf O(-D))_\mathbf{r}$ can be nonzero
only if the divisor $\sum r_iD_i$ is homologous to $-D$.  An obvious way to
satisfy this necessary condition is if $r_i=-1$ for all $i$.  We put
$\mathbf{-1}=(-1,\ldots,-1)$.  Since $H^n(X,\sheaf O(-D))$ is one dimensional, our
conclusion is that

\medskip
\begin{center}
{\it If $H^n(X,\sheaf O(-D))_\mathbf{-1}\ne0$, then $\dim H^n(X,\sheaf
O(-D))_\mathbf{-1} = 1$ and\\
$H^n(X,\sheaf O(-D))_\mathbf{-1}\simeq H^n(X,\sheaf O(-D))\simeq\CC$.}
\end{center}
\medskip

This means that if we can verify that the group $H^{\rm top} (\sheaf
M,\sheaf O(-D))_\mathbf{-1}$ is nonzero for the instanton moduli space $\sheaf M$, then
it is one dimensional, and we only have to consider \v Cech cocycles whose
components are multiples of $1/(x_0\cdots x_N)$ in the computation of \v Cech
cohomology and the determination of a trace map.  We conjecture that this
condition always holds, a conjecture implicitly made in developing the
algorithm in \cite{Katz:2004nn}.\\

For the purposes of computing the cohomology, not every intersection
$U=U_{\sigma_0}\cap\ldots\cap U_{\sigma_k}$ is relevant, since $1/(x_0\cdots
x_N)$ is required to be holomorphic there.  In terms of the fan, the condition
is that for each edge $\rho$, there must be at least one top-dimensional cone
$\sigma_i$ used in indexing $U$ which does not contain $\rho$.  We call these
open sets $U$ {\em good\/} open sets.\\

Thus, when taking cup\slash wedge products to obtain \v Cech $n$-cochain
representatives of correlation functions, we do not account for their value on
every $n+1$-fold intersection of open sets.  For example, on $\pones$, there
are four open sets: $\uu 00$, $\uu 10$, $\uu 01$, and $\uu 11$, with $\uu ij$
defined as in \eqref{eq:pmnOpenCover}.  The ``good'' two-fold intersections
would be $\uc 0011$, and $\uc 1001$.  The intersection $\uc 0010$ would be
excluded, for example, since $y_1$ is not invertible on this set.

\subsection{The Trace Map}

For any given instanton sector, we have $H^\text{top} (\sheaf F^*) \cong \CC$,
as follows from \eqref{eq:anomcan}.  In order to complete our computations, we
will need the explicit form of this isomorphism,  i.e.\ a trace map. For the
reasons discussed in the previous section, we can and will replace
$H^\text{top} (\sheaf F^*)$ with $H^\text{top} (\sheaf F^*)_\mathbf{-1}$.
Since we will be dealing with cup\slash wedge products of elements of
\(C^1 \Big (\sheaf O(-1,0)^{2m+2} \oplus \sheaf O(0,-1)^{2n+2}\Big )\)
rather than explicit elements of $H^\text{top}$, we will actually construct a
map from $C^\text{top}(K_{\sheaf M})_\mathbf{-1}$ to $\CC$ and use the minor
map to remove the extraneous terms introduced by using cochains rather than
cocycles to compute.\\

The desired map is determined up to a non-zero complex multiple by the
following algorithm. The $\delta$-map between cochain groups can be described
by specifying the weight with which the components of an element of $C^{i-1}$
on each $i$-fold intersection contribute to the components on each $(i+1)$-fold
intersection in $C^i$. For any element $A$ in $C^{i-1}$, and on each
$(i+1)$-fold intersection in $C^i$, we obtain the representative of $\delta A$
on that intersection by summing elements in $C^{i-1}$:
\[
(\delta A)_{j_1,\cdots,j_{i+1}} = \sum_{k = 1}^{i+1} (-1)^{k-1}
A_{j_1,\cdots,\hat \jmath_k,\cdots,j_{i+1}}.
\]

In this expression, $\hat \jmath_k$ denotes the exclusion of the
$j_k^{\,\text{th}}$ index from the collection.  Writing elements of $C^i$ as a
column vector, with each slot in the vector denoting the element's value on an
$(i+1)$-fold intersection, the $\delta$ map is simply a matrix with elements
$0$, $1$, or $-1$.  Then it is straightforward to obtain $\delta$:  if the
$i$-fold intersection denoted by $j_1,\cdots,\hat \jmath_k,\cdots,j_{i+1}$ is
in the $n^\text{th}$ row of the vector, and the $(i+1)$-fold intersection
$j_1,\cdots,j_{i+1}$  is in the $m^\text{th}$ row, then the $(m,n)^\text{th}$
element of $\delta_{i-1}$ is $(-1)^k$.  Let us reiterate: since we are
interested in cohomology, and invalid intersections will not contribute, we can
ignore them when computing $\delta$.\\

Next, we construct a matrix $Z^i$, whose rows span the nullspace of $\delta_i$.
We consider the matrix $Z^i$ as a projection map from $C^i$ to the basis of
$\ker(\delta_i)$ given by the rows of $Z^i$.     In addition, if we multiply
this matrix by its transpose, we obtain a projection map from $C^i$ to
$\ker(\delta_i) \subset C^i$, expressed in the basis for $C^i$: by composing
$\delta_{i-1}$ with ${}^\text tZ^i \cdot Z^i$, we obtain a map 
\begin{equation}
  ({}^\text tZ^i \cdot Z^i) \circ \delta_{i-1} \takes
  {C^{i-1}}{\ker(\delta_i)\subset C^i}.
  \label{eq:toKernel}
\end{equation}

The transpose of $Z^i\circ \delta_{i-1}$ is a map from $\ker(\delta_i)$ to
$C^{i-1}$, whose kernel $H_i$ spans the subspace of elements of
$\ker(\delta_i)$ which are not in the image of $\delta_{i-1}$: the row space of
this matrix is exactly the $i^\text{th}$ cohomology group we seek,
\begin{equation}
  \begin{split}
	 H_i &= \ker \left[ {}^\text t(Z^i \circ \delta_{i-1})\right]\\
	 H^i &\equiv \text{rowspace}(H_i).
  \end{split}
  \label{eq:cohom}
\end{equation}

In the same manner as $Z^i$, we think of $H_i$ as a map from $\ker(\delta_i)$
to $H^i$.  Thus, the composition $H_i\cdot Z^i$ is a projection from $C^i$ to
$H^i$, the desired trace map.\\

Mathematica\cite{Mathematica:52} was sufficient for computing $\delta$ kernels for most instanton
sectors.  However, for the degree $(1,1)$ sector, the cocycle space's large
dimension required the use of specialized software to find null spaces.  We
used routines from the ``Integer Matrix Library''
\cite{Chen:2005sj,Chen:2005ts} to compute bases for various kernels over
$\ZZ_p$ for $p = 5279, 4409, 3571$, before applying the Chinese Remainder
Theorem to obtain bases over the integers.  For a map $\delta_i$, the result is
the matrix form of $Z^i$; its rows form a basis for $\ker(\delta_i)$.

\subsection{Applicability}

The algorithm described in this section may be utilized for any compact toric target
space whose fan's support is a cone and is simplicial.  See \cite{Morrison:1994fr} for
generalities on toric varieties and their relationship to the GLSM.\\

Given a general GLSM containing bosonic fields $\phi_i, i=0\ldots N$ with
respective charges $Q_{ik},\ k=1..r$ under a $U(1)^r$ gauge group, we denote by
$M$ the vacuum moduli space in a geometric phase. Geometrically, $M$ can be
described as a toric variety with edges $v_i$ in one to one correspondence with
the $\phi_i$.  Each $v_i$ is associated to a divisor $D_i$ on $M$, along with
a line bundle $\sheaf O(D_i)$.  There are sections $x_i$ of the
bundle $\sheaf O(D_i)$ that serve as homogeneous coordinates on $M$, from
which $M$ can be recovered from $\CC^{N+1}$ by symplectic reduction in the usual
way.\\

There is an exact sequence, proven in
\cite{Batyrev:1994}, which generalizes \eqref{eq:ponesles} for any
quasi-smooth compact toric variety $M$:
\begin{equation}
\begin{diagram}
  0 \lra
  \sheaf O^r &
  \lra \oplus_i \sheaf O(D_i)
 \lra T_M \lra 0 .
\end{diagram}
\label{eq:toricles}
\end{equation}

The map $\oplus\sheaf O(D_i)\to T_M$ takes a collection of sections $s_i$ of
$\sheaf O(D_i)$ to the tangent vector $\sum_i s_i(\partial/\partial x_i)$,
which is a well-defined vector field on $M$ since it is neutral under $U(1)^r$
by construction.  The $(i,k)^\text{th}$ entry of the map $\oplus\sheaf O^r\to \oplus_i
\sheaf O(D_i)$ is given by $Q_{ik}x_i$.  Note that
$\sum_iQ_{ik}x_i(\partial/\partial x_i)$ is the Euler vector field in
$\CC^{N+1}$
corresponding to the action of the $k^{\scriptstyle \mathrm{th}}$ $U(1)$ on
$\CC^{N+1}$ that gives rise to the trivial vector field on $M$, which is one of the
requirements of the exactness of \eqref{eq:toricles}. \\

The tangent bundle $T_M$ can be deformed as in the case of $M=\pones$ by
deforming the entries of the first non-trivial map in \eqref{eq:toricles},
giving an exact sequence
\begin{equation}
\begin{diagram}
  0 \lra
  \sheaf O^r &
  \lra \oplus_i \sheaf O(D_i)
 \lra \sheaf E \lra 0.
\end{diagram}
\label{eq:deformedles}
\end{equation}

Dualizing, we have an exact sequence
\begin{equation}
\begin{diagram}
  0 \lra
  \sheaf E^* &
  \lra \oplus_i \sheaf O(-D_i)
 \lra \sheaf O^r \lra 0.
\end{diagram}
\label{eq:dualses}
\end{equation}

The coboundary map in (\ref{eq:dualses}) is $H^0(\sheaf O^r)\to H^1(\sheaf
E^*)$, which can be used to describe elements of $H^1(\sheaf E^*)$ as in the
$\pones$ case, by taking lifts.  While the lifts cannot be written in closed
form as simply as in the $\pones$ case, they can be found algorithmically.\\

In an instanton sector, we get a sheaf $\sheaf F$ on the instanton moduli space
$\sheaf M$.  As described in \cite[Section 3.7]{Morrison:1994fr}, $\sheaf M$ is
itself a toric variety.  To each edge $v_i$ in the fan for $M$ is associated
$b_i$ edges in the fan for $\sheaf M$, with the number $b_i$ depending on the
particular instanton sector under consideration.  Since $\sheaf M$ is a toric
variety, we have analogous to (\ref{eq:dualses}) a short exact sequence
\begin{equation}
\begin{diagram}
  0 \lra
  \sheaf F^* &
  \lra \oplus_i \sheaf O(-D_i)^{b_i}
 \lra \sheaf O^r \lra 0.
\end{diagram}
\label{eq:dualinstles}
\end{equation}
generalizing (\ref{eq:modulises}).  Here we are abusing notation by denoting
with $D_i$ the divisor class on $\sheaf M$ associated with the divisor class
$D_i$ on $M$ as described in \cite{Morrison:1994fr}.  Taking determinants in
(\ref{eq:dualinstles}) gives $\det(\Lambda^{\rm top}F^*)\simeq \sheaf O(-\sum_i
b_iD_i)\simeq K_{\sheaf M}$ as before.\\

The coboundary map of (\ref{eq:dualinstles}) is $H^0(\sheaf O^r)\to H^1(\sheaf
F^*)$.  Since the domain of this map is naturally $\CC^r$, as is the domain of
the other coboundary map $H^0(\sheaf O^r)\to H^1(\sheaf E^*)$, we have a
natural way to map our elements of $H^1(\sheaf E^*)$ to elements of $H^1(\sheaf
F^*)$.\\

To compute correlation functions of our elements of $H^1(\sheaf E^*)$, we map
them to $H^1(\sheaf F^*)$ as just explained and take cup products to get an
element of $H^{\rm top}(K_{\sheaf M})\simeq\CC$.  The methods of
\cite{Katz:2004nn} together with the methods developed here for computing the
trace map allow for the completion of the desired computation in principle,
constrained only by the limitations of machine computation.

\section{Chiral Ring Relations}
\label{sec:cohomologyRelations}

\subsection{Calculation}
\label{ssec:qcohomCalculation}

The quantum cohomology ring is the polynomial ring $\CC[\psi,\wt \psi]$,
modulo two relations, which in the $(2,2)$ limit reduce to 
\begin{equation}
  \begin{split}
  \psi \star \psi &= q \\
  \wt \psi \star \wt \psi &= \wt q.
  \end{split}
  \label{eq:22ringreltn}
\end{equation}

We can partially deduce the form of the relations by making the ans\"atze
\begin{equation}
  \begin{split}
  a\, (\psi \star \psi) + b \, (\psi \star \wt \psi) + c\,  (\wt \psi \star \wt \psi) &= d\\
  \wt a\, (\psi \star \psi) + \wt b \, (\psi \star \wt \psi) + \wt c\,  (\wt \psi \star \wt \psi) &= \wt d,
  \end{split}
  \label{eq:ringreltn}
\end{equation}
where for example $\wt a\rightarrow 0$ in the $(2,2)$ limit.  The problem
possesses an inherit symmetry, wherein
\begin{equation}
  \epsilon_i \leftrightarrow \gamma_i \qquad\qquad\qquad
  \psi       \leftrightarrow \wt \psi \qquad\qquad\qquad
  x^i        \leftrightarrow  y^i \qquad\qquad\qquad
  q          \leftrightarrow   \wt q,
  \label{eq:probsymmetryParams}
\end{equation}
which implies that the first and second ring relations in \eqref{eq:ringreltn}
are exchanged under the symmetry with the identifications
\begin{equation}
  a \leftrightarrow \wt c \qquad\qquad\qquad
  b \leftrightarrow \wt b \qquad\qquad\qquad
  c \leftrightarrow \wt a \qquad\qquad\qquad 
  d \leftrightarrow \wt d.
  \label{eq:probsymmetryCoeff}
\end{equation}

Thus, only one ring relation need be computed.  The relations may be deduced by
identifying cohomological inner product and correlation functions in the
physical theory:
\begin{equation*}
  (\psi \star \cdots \star \wt \psi, \psi) = \left \langle \psi \cdots \wt \psi  \psi \right \rangle.
\end{equation*}
Then, using the commutativity of the ring, correlators are related by
\eqref{eq:ringreltn}.  For example, since $d$ is a constant, we will have that
$\langle d \rangle = 0$ by anomaly considerations, and the two-point functions
may be related as 
\begin{equation}
  a \left\langle \psi \psi\vphantom{\wt \psi} \right\rangle  + b \left\langle \psi \wt \psi \right\rangle + c
  \left\langle \wt \psi \wt \psi \right\rangle = 0.
  \label{eq:relation1}
\end{equation}
In general, we will have 
\begin{equation}
  a \left\langle \cdots \psi \psi \psi \wt \psi \cdots \vphantom{\wt \psi} \right\rangle +
  b \left\langle \cdots \psi \psi \wt \psi \wt \psi \cdots \right\rangle +
  c \left\langle \cdots \psi \wt \psi \wt \psi \wt \psi \cdots  \right\rangle = 
  d \left\langle \vphantom{\wt \psi} \cdots \psi \wt \psi \cdots \right\rangle
  \label{eq:relation2}
\end{equation}
where ``$\cdots$'' on the left side of the correlator is an equal, but
arbitrary, number of $\psi$'s, and similarly for $\wt \psi$'s on the right.
Each computed correlation function (up to instanton degree 2) was consistent,
but only 2 were independent.  Thus, we deduce the constants in the ring
relation \eqref{eq:ringreltn} by substitution into some of these four-point
functions.
\begin{align*}
  a  \left \langle \vphantom{\wt \psi} \psi\psi\psi\psi \right \rangle &=
  -b \left \langle \psi\psi\psi\wt\psi \right \rangle 
  -c \left \langle \psi\psi\wt\psi\wt\psi \right \rangle 
  +d \left \langle \vphantom{\wt \psi}\psi\psi \right \rangle \\
  a  \left \langle \psi\psi\psi\wt\psi \right \rangle &=
  -b \left \langle \psi\psi\wt\psi\wt\psi \right \rangle 
  -c \left \langle \psi\wt\psi\wt\psi\wt\psi \right \rangle 
  +d \left \langle \psi\wt\psi \right \rangle\\ 
  a  \left \langle \psi\psi\wt\psi\wt\psi \right \rangle &=
  -b \left \langle \psi\wt\psi\wt\psi\wt\psi \right \rangle 
  -c \left \langle \wt\psi\wt\psi\wt\psi\wt\psi \right \rangle 
  +d \left \langle \wt\psi\wt\psi \right \rangle 
\end{align*}

Plugging in the values for the correlation functions appearing in Appendix
\ref{sec:correlators} and solving these relations, we determine that the
coefficients are related by the equations
\begin{align*}
  b \left(1-\gamma _2 \gamma _3 \epsilon _2 \epsilon _3\right)
  &=\left(a+c \gamma _2 \gamma _3\right) \epsilon _1+\gamma _1 \left(c+a \epsilon _2 \epsilon _3\right)\\
  d \left(1-\gamma _2 \gamma _3 \epsilon _2 \epsilon _3\right)
  &=c \left(\wt q +q \gamma _2 \gamma _3\right)+a \left(q+\wt q \epsilon _2
  \epsilon _3\right).
\end{align*}

In the (2,2) limit, the ring relation constants must reduce to 
\[ a = 1 \qquad\qquad b = 0 \qquad\qquad c = 0 \qquad\qquad d = q,\]

so we make the substitutions $a \rightarrow 1 + a$ and $d \rightarrow q +
d$ and the equations become
\begin{align*}
  b \left(1-\gamma _2 \gamma _3 \epsilon _2 \epsilon _3\right)
  &=\left[ (1 + a) +c \gamma _2 \gamma _3\right] \epsilon _1+\gamma _1 \left[c+(1+ a) \epsilon _2 \epsilon _3\right]\\
  (q + d) \left(1-\gamma _2 \gamma _3 \epsilon _2 \epsilon _3\right)
  &=c \left(\wt q +q \gamma _2 \gamma _3\right)+ (1 + a) \left(q+\wt q \epsilon _2 \epsilon _3\right),
\end{align*}

where now $a,b,c,$ and $d$ all vanish in the (2,2) limit.  We will assume that
these constants are independent of $q$ and $\wt q$, and will provide arguments
partially justifying this assumption at the end 
of \S \ref{ssec:mirrorSymmetryComputation}. 
We thus collect the coefficients of $q$ and $\wt q$, set them equal to
zero, and solve for $\{a,b,c,d\}$.  This procedure produces an
astonishingly simple relation: coupled with the symmetry described in
\eqref{eq:probsymmetryCoeff}, we have that
\begin{alignat}{4}
  \label{eq:solvedRingRelation1}
  \psi *\psi                          & +\epsilon _1 (\psi  *\wt \psi ) && {}-\epsilon _2 \epsilon _3 &&(\wt \psi *\wt \psi ) && =q\\
  \wt{\psi }*\wt{\psi } & +\gamma _1 (\psi *\wt \psi )    && {}-\gamma _2 \gamma _3 &&(\psi  *\psi)                           && =\wt{q}.
  \label{eq:solvedRingRelation2}
\end{alignat}

\subsection{Further Evidence}

Here, we offer some further evidence for our ring relations.  Starting with the
GLSM description of the $(0,2)$ theory as outlined in \S
\ref{sec:classicalOperators}, one would suspect that the classical ring
relations (those with $q = 0$) might be encoded into the classical action
somehow.  We investigate the Coulomb branch of the GLSM in search of these
relations.\\

The action for the left-moving fermions contains the term $\sum_i \abs
{E_i}^2$.  Here, $E_i$ is the function in \eqref{eq:originOfE}, which upon
performing the superspace integrals depends only on the lowest components of
the superfields.  After integrating out the $D$ field, the bosonic interactions
that remain arise from this expression.  Writing these terms in matrix form,
we have 
\begin{equation}
  \left (\begin{matrix} \bar \phi_1 \\ \bar \phi_2 \end{matrix}\right )
  \left(
  \begin{array}{ll}
         \abs{\Sigma }^2 + \epsilon_1 \wt{\Sigma} \bar{\Sigma}+ \bar{\epsilon_1} \Sigma \bar{\wt{\Sigma}}+ \abs{\epsilon_1\wt{\Sigma}}+ \abs{\epsilon_3\wt{\Sigma}}^2
         & \bar{\epsilon_3} \Sigma \bar{\wt{\Sigma}}+
         \bar{\epsilon_1} \epsilon_2 \abs{\wt{\Sigma}}^2 +
          \epsilon_2\bar{\Sigma} \wt{\Sigma} \\
        \bar{\epsilon_2} \Sigma  \bar{\wt{\Sigma}}+
        \epsilon_1 \bar{\epsilon_2} \abs{\wt{\Sigma}}^2 +
        \epsilon_3 \bar{\Sigma} \wt{\Sigma} & 
          \abs{\Sigma }^2 + \abs{\epsilon_2\wt{\Sigma}}^2 
  \end{array}
  \right )
  \left (\begin{matrix} \phi_1 \\ \phi_2 \end{matrix}\right )
  \label{eq:bosonicMassTerm1}
\end{equation}
and
\begin{equation}
  \left (\begin{matrix} \overline {\wt \phi}_1 \\ \overline {\wt \phi}_2\end{matrix}\right )
  \left(
  \begin{array}{ll}
         \abs{\wt{\Sigma}}^2
         + \gamma_1\Sigma  \bar{\wt{\Sigma}}
        + \bar{\gamma_1} \bar{\Sigma} \wt{\Sigma}
         + \abs{\gamma_1\Sigma}^2 
         + \abs{\gamma_3 \Sigma}^2 &
          \bar{\gamma_3} \bar{\Sigma} 
          \wt{\Sigma}+\bar{\gamma_1} \gamma_2\abs{\Sigma}^2 +\gamma_2 \Sigma \bar{\wt{\Sigma}} \\
        \bar{\gamma_2} \bar{\Sigma} 
        \wt{\Sigma}+
        \gamma_1\bar{\gamma_2} \abs{\Sigma}^2
        +\gamma_3 \Sigma 
          \bar{\wt{\Sigma}} &
          \abs{\wt{\Sigma}}^2 + \abs{\gamma_2\Sigma}^2
  \end{array}
  \right)
  \left (\begin{matrix} \wt \phi_1 \\ \wt \phi_2 \end{matrix}\right ).
  \label{eq:bosonicMassTerm2}
\end{equation}

Let us concentrate in particular on the contribution of the bosonic zero modes
to the $\Sigma$ effective action.   From the matrix forms
\eqref{eq:bosonicMassTerm1} and \eqref{eq:bosonicMassTerm2}, one can read off
that the zero modes' contribution  appears in the combination
\begin{equation} 
        \left\vert
        \tilde{\Sigma}^2 + \gamma_1 \Sigma \tilde{\Sigma} -\gamma_2 \gamma_3 \Sigma^2
        \right\vert^2
        \left\vert
        \Sigma ^2+
   \epsilon_1 \Sigma \tilde{\Sigma}-\epsilon_2 \epsilon_3 \tilde{\Sigma}^2 
        \right\vert^2.
\end{equation}
These bosonic modes precisely cancel the fermionic zero mode contribution,
which arise from Yukawa interactions of the form
\begin{equation}
  \sigma \bar \rho_a \frac{\del E^a}{\del \phi_i} \psi_i + \text{c.c.},
  \label{eq:yukawaInteractions}
\end{equation}
where $\sigma$ is the lowest component of the superfield $\Sigma$.  Note that
the multiplicands are exactly the classical limit of the relations in
\eqref{eq:solvedRingRelation1} and \eqref{eq:solvedRingRelation2}.  These
results provide strong indications that our ring relations are correct.

\section{Comparison with Previous Results}
\label{sec:comparison}

Two questions naturally arise: do the chiral ring relations
\eqref{eq:solvedRingRelation1} and \eqref{eq:solvedRingRelation2} match the
results in \cite{Adams:2003zy}, and do the correlation functions outlined in
\eqref{eq:degree00correlators}--\eqref{eq:degree01correlators} match those in
\cite{Katz:2004nn}?  Let us first address the latter question.

\subsection{Cohomological Computations}

The deformation described in \cite{Katz:2004nn} relied on two parameters,
$\epsilon_1$ and $\epsilon_2$,  yet the ring relations produced therein
depended only on the difference of these two parameters.  This dependence
arises from the fact that deformations described by $\epsilon_2$ are not
independent from those described by $\epsilon_1$:  consider their matrix,
\begin{equation}
  \left (
  \begin{matrix}
         x_0 & x_1 & 0 & 0 \\
         \epsilon_1 x_0 & \epsilon_2 x_1 & y_1 & y_2
  \end{matrix}
  \right ).
  \label{eq:ksMatrix}
\end{equation}
By adding arbitrary multiples of the first row to the second, one generates the
same translation of both parameters.  Since both describe the same deformation,
we set the $\epsilon_2$ in \cite{Katz:2004nn} to zero so that the deformation
is fully described by $\epsilon_1$.\\

In order to match our results, the $\epsilon$ and $\gamma$ parameters must be
adjusted so that the deformation described by
\eqref{eq:generalDeformationMatrix} matches that of \cite{Katz:2004nn}, under
the condition on their parameters imposed above.  In particular, the
deformations match when 
\[\epsilon_2 = \epsilon_3 = \gamma_1 = \gamma_2 = \gamma_3 = 0.\]
In this limit, one can check that the four-point functions appearing in \S 7.1
of \cite{Katz:2004nn} match those in Appendix \ref{sec:correlators}.  The
results of our six-point computations do not appear due to their size, but they
do match the results of \cite{Katz:2004nn}, as well as being consistent with
the relations \eqref{eq:solvedRingRelation1} and
\eqref{eq:solvedRingRelation2}.\\

\subsection{Computation via Mirror Symmetry}
\label{ssec:mirrorSymmetryComputation}

We now begin our comparison with the results of \cite{Adams:2003zy}.  Their
work centered on the construction of dual theories, and the exploitation of
this symmetry to deduce the form of the chiral ring relations, among other
properties.  In \S 6.2, the authors construct a GLSM coupled to a deformation
of the tangent bundle of $\pones$.\\

The fermi superfields in the sigma model are not quite chiral, with their
deviation from chirality measured by combinations $E_i$ and $\wt E_i$ of chiral
fields $\Phi_i,\wt \Phi_i, \Sigma$ and $\wt \Sigma$.  These functions encode
the deformation matrix in a straightforward way; the $E$'s may be interpreted
as the image in $\Gamma \big(\sheaf O(1,0)^2 \oplus \sheaf
O(0,1)^2\big)$ of the basis elements $\Sigma$ and $\wt \Sigma$ by the injection
in the short exact sequence \eqref{eq:ses}.\\

Compare the functions of chiral fields in \S 6.2 with our deformation mapping
\eqref{eq:generalDeformationMatrix}.  The parameters $\alpha_2^\prime$ and
$\beta_2^\prime$ are dependent on the other $\alpha$'s and $\beta$'s, since the
space of deformations is only six-dimensional: the first of these variables
corresponds to $\epsilon_2$ of \cite{Katz:2004nn}.  We have argued previously
that $\epsilon_2$ does not measure an independent deformation, and under the
symmetry described in \eqref{eq:probsymmetryParams}, $\alpha_2^\prime$ and
$\beta_2^\prime$ are exchanged. Upon setting these extraneous parameters to
zero, we find agreement to our deformation when
\begin{equation}
\alpha_i = \epsilon_i
\qquad
\alpha^\prime_1 = \epsilon_3
\qquad
\beta_i = \gamma_i
\qquad
\beta_1^\prime = \gamma_3.
  \label{eq:absConversion}
\end{equation}

Chiral ring relations were computed by comparing effective potentials for those
fields unaffected by duality in both the GLSM and its mirror dual, resulting in 
\begin{equation}
  \begin{split}
         X + p(\epsilon, \gamma) \frac {e^{it_1}} X + q(\epsilon, \gamma) \wt X + s(\epsilon,\gamma) \frac {e^{it_2}} {\wt X} &= 0\\
         \wt X + \wt p(\epsilon, \gamma) \frac {e^{it_2}} {\wt X} + \wt q(\epsilon, \gamma) X + \wt s(\epsilon,\gamma) \frac {e^{it_1}} X &= 0.
  \end{split}
  \label{eq:absrelation}
\end{equation}

Here, $e^{it_1}$  corresponds to our $q$ and $e^{it_2}$ to our $\wt q$.  The
chiral fields $X$ and $\wt X$, as elements of the quantum cohomology ring, must
be related by some simple linear combination to our operators $\psi$ and $\wt
\psi$,
\begin{equation}
  \left (\begin{matrix} X \\ \wt X \end{matrix}\right) = 
  \left (\begin{matrix} A & B \\ C & D \end{matrix}\right)
  \left (\begin{matrix} \psi \\ \wt \psi \end{matrix}\right)
  \label{eq:changeOfVariables},
\end{equation}

where each element of the matrix is some polynomial in the $\epsilon$'s and
$\gamma$'s.  Assuming that the ring relation \eqref{eq:solvedRingRelation1}
will take the form \( \alpha X^2 + \beta X \wt X + \gamma \wt X^2 = q,\) we
plug in the change of variables above and find that 
\begin{equation}
  \alpha,\beta, \gamma \propto \frac 1 {(B C-A D)^2}.
  \label{eq:coeffProportional}
\end{equation}

At first glance this is discouraging, since the ring relation expected from the
physics seems to be ill-defined on certain configurations of parameters.  All
is not lost, however: we have encountered this type of situation before.  To
alleviate our conceptual difficulties, we make the ansatz that the only
combination of the deformation parameters that could possibly appear in the
denominator of such a change of variables is the polynomial $\phi$ defined in
\eqref{eq:phiDef}:
\[BC - AD = \phi.\]

Let us now impose our knowledge of the behavior of \eqref{eq:changeOfVariables}
in the limiting case examined in \cite{Katz:2004nn}.  There, it was found that
\( X = \psi + \epsilon_1 \wt \psi\), \( \wt X = \wt \psi.\) Notice that due to
symmetry, had we chosen $\gamma_1$ to be the non-vanishing parameter, the
correct change of variables would have been \( X = \psi\), \(\wt X =\gamma_1
\psi + \wt \psi.\) With this observation, we make the ansatz that the correct
limiting transformation when $\epsilon_1$ and $\gamma_1$ are non zero is 
\[ X = \psi + \epsilon_1 \wt \psi \qquad\qquad \wt X =\gamma_1 \psi + \wt \psi.\]

We make a change of variables to account for this limiting behavior:
\begin{align*}
  A &\to 1 + A^\prime &
  B &\to \epsilon_1 + B^\prime\\ 
  C &\to 1 + C^\prime & 
  D &\to \gamma_1 + D^\prime.
\end{align*}

In \cite{Adams:2003zy}, a charge was assigned to each parameter:  in that
scheme, $\epsilon_1$ was given charge $k - \wt k$, while $\gamma_1$ was given
charge $\wt k - k$.  In order to match the charges of the limiting behaviour,
$A^\prime$ and $D^\prime$ must be neutral, while $B^\prime$ and $C^\prime$ must
have charges $k - \wt k$ and $\wt k - k$, respectively.  Analyticity of the
basis change requires the coefficients to be polynomials in the parameters, and
relevant combinations of parameters and charges appear in Table
\ref{table:charges}.    
\begin{table}[ht]
  \begin{tabular}{|c|c|}
         \hline
         $k - \wt k \tb$ & $ \wt k - k$ \\
         \hline
         $\epsilon_1\tb$ & $\gamma_1$\\
         $\gamma_1 \epsilon_2 \epsilon_3$ & $\epsilon_1 \gamma_2
         \gamma_3\tb$ \\
         \hline
  \end{tabular}
  \vspace*{5pt}
  \caption{Charge assignments for relevant combinations of deformation parameters.}
  \label{table:charges}
\end{table}

Since we have already accounted for the $\epsilon_1$ dependence of $B$, the
only other possibility is $\gamma_1 \epsilon_2 \epsilon_3$.  We make the
ans\"atze that $B = \epsilon_1 + \gamma_1 \epsilon_2 \epsilon_3$, and $C =
\gamma_1 + \epsilon_1 \gamma_2 \gamma_3$, so that upon substitution \(\phi = B
C - A D \) reduces to 
\begin{equation}
  0 = -(\gamma_2 \gamma_3 \epsilon_2 \epsilon_3)^2+2 \gamma_2 \gamma_3
  \epsilon_2 \epsilon_3+A^\prime+A^\prime D^\prime+D^\prime.
  \label{eq:ADEqn}
\end{equation}

One observes that this equation is solved by \( A^\prime = D^\prime = -\gamma_2
\gamma_3 \epsilon_2 \epsilon_3,\) a combination of charge zero, so that the
proper change of variables is 
\begin{equation}
  \left (\begin{matrix} X \\ \wt X \end{matrix}\right) = 
  \left (
  \begin{matrix}
         1 - \gamma_2 \gamma_3 \epsilon_2 \epsilon_3 &
         \epsilon_1 + \gamma_1 \epsilon_2 \epsilon_3 \\
         \gamma_1 + \epsilon_1 \gamma_2 \gamma_3 & 
         1 - \gamma_2 \gamma_3 \epsilon_2 \epsilon_3 
  \end{matrix}
  \right)
  \left (\begin{matrix} \psi \\ \wt \psi \end{matrix}\right)
  \label{eq:fullChangeOfVariables}.
\end{equation}
This change of variables respects the interchange symmetry described by
\eqref{eq:probsymmetryParams};  exchange of the $\psi$'s and the $\epsilon$ and
$\gamma$ parameters results in an exchange of the $X$'s.\\

With this change, the coefficients in the ring relation for the new variables
are easily computed to be
\begin{equation}
 \left . 
 \begin{array}{rl}
        \alpha =& \!\!\!\!-\frac 1 \phi \\
        \beta  =& \!\!\!\!\frac 1 \phi (\epsilon_1)  \\ 
        \gamma =& \!\!\!\!\frac 1 \phi (\epsilon_2 \epsilon_3) 
  \end{array} 
\right \}
\Rightarrow
\begin{matrix}
- X^2 + \epsilon_1 X \wt X + \epsilon_2 \epsilon_3 \wt X^2 = \phi \; q \\ 
- \wt X^2 + \gamma_1 X \wt X + \gamma_2 \gamma_3  X^2 = \phi \; \wt q \\ 
\end{matrix}\;.
  \label{eq:covRelations}
\end{equation}

Let us now turn to a comparison of these results with those
\eqref{eq:absrelation} deduced in \cite{Adams:2003zy}.  We find an
apparent contradiction between our ring relation in the new variables
\eqref{eq:covRelations} and those in \eqref{eq:absrelation}.  We begin by
multiplying  the equations in \eqref{eq:absrelation} by $X \wt X$,
obtaining
\begin{equation}
  \begin{split}
         X^2 \wt X +  p(\epsilon, \gamma) \wt X e^{it_1} + q(\epsilon, \gamma) X \wt
         X^2 + s(\epsilon,\gamma) X e^{it_2} &= 0\\
         X\wt X^2 + \wt p(\epsilon, \gamma) X e^{it_2} + \wt q(\epsilon,
         \gamma) X^2 \wt X + \wt s(\epsilon,\gamma) \wt X  e^{it_1} &= 0.
  \end{split}
  \label{eq:absrelationMultiplied}
\end{equation}

One can quickly observe that there is no linear combination of these two
relations which would give those in \eqref{eq:covRelations}, since there
would need to be terms like $X^3$ and $\wt X^3$ in
\eqref{eq:absrelationMultiplied}.\\

Let us consider this problem from another direction; substituting $\psi$ and
$\wt \psi$ into the relations \eqref{eq:absrelationMultiplied}.  We impose that
upon restriction to the $(2,2)$ locus, the first (second) ring relation matches
the A-model relation $X^2 = q$ ($\wt X^2 = \wt q$).  Then, $s(\epsilon,\gamma)
= \wt s(\epsilon,\gamma) = 0$, and up to multiplication by some other function
of the $\epsilon$'s and $\gamma$'s, the relations become
\begin{equation}
  \begin{split}
  X^2 + q(\epsilon,\gamma) X \wt X &= -p(\epsilon,\gamma) e^{it_1}\\
  \wt X^2 + \wt q(\epsilon,\gamma) X \wt X &= -\wt p(\epsilon,\gamma) e^{it_2}.
  \end{split}
  \label{eq:generalRingRelation}
\end{equation}

Then, we make the change of variables
\begin{equation}
  \left ( \begin{matrix} \psi \\ \wt \psi \end{matrix} \right ) =
  \left ( \begin{matrix} \wt A & \wt B \\ \wt C & \wt D \end{matrix} \right )
  \left (\begin{matrix} X \\ \wt X \end{matrix} \right ),
  \label{eq:changeOfVariables2}
\end{equation}

and substitute into the ring relations \eqref{eq:solvedRingRelation1} and
\eqref{eq:solvedRingRelation2}.  Matching the $X$ and $\wt X$ dependence in
\eqref{eq:generalRingRelation}, we must have that 
\begin{equation}
  \wt B = - \frac{\wt D}{2} \left( \epsilon_1 \pm \sqrt{\epsilon_1^2+4 \epsilon_2
  \epsilon_3}   \right )
  \qquad
  \wt C =  - \frac{\wt A}{2} \left(\gamma_1 \pm \sqrt{\gamma _1^2+4 \gamma _2 \gamma
  _3} \right).
\end{equation}

Thus, we find that the functions $q(\epsilon,\gamma)$ and $p(\epsilon,\gamma)$
must depend on square roots of combinations of the parameters, which appear
in both the numerator and denominator of these functions.  Unfortunately, it is not possible
to completely solve for $\{\wt A,\wt B,\wt C,\wt D\}$, since there are four unknowns
and only two equations imposed by $X$ and $\wt X$ dependence after
substitution into \eqref{eq:generalRingRelation}.
It may be that this is the correct change of variables, although it is not
obvious that it is well defined when the bundle is, and is 
\ae sthetically quite displeasing besides.  \\


It may be possible, however, to resolve the apparent contradiction between
the ``nice'' change of variables \eqref{eq:covRelations} and the results of
\cite{Adams:2003zy}.  The ring relations therein were computed by simply
minimizing the superpotential in the theory dual to the GLSM.  It may be
that this is only true classically (in the field theory sense).  Since the
ratio of left- and right-moving determinants need not cancel, one may find
left-moving one loop corrections.  Indeed,  some preliminary work\cite{Sethi:2007pv} indicates
that the appearance of terms quadratic in the deformation parameters
qualitatively matches expectations of one-loop
contributions.\\

Additionally, note that the degree $(0,0)$ correlators appearing in
\eqref{eq:degree00correlators} have terms which are cubic and quartic in the
deformation parameters.  Given the form of the interactions in the GLSM --
outlined in equations \eqref{eq:bosonicMassTerm1}, \eqref{eq:bosonicMassTerm2},
and \eqref{eq:yukawaInteractions} -- such terms would most likely arise from loop
contributions.\\

Let us say a few more words about the coefficients of \eqref{eq:ringreltn}.  We
first note that the chiral ring relations in  \eqref{eq:absrelation} are
independent of $q$ and $\wt q$.  The Yukawa couplings contributing to
possible one-loop additions to the dual superpotential do not involve $q$ or
$\wt q$ either, so that from a physical standpoint, the coefficients cannot
develop dependence on these quantities.

Consider this from a the point of view of the charges of the parameters, as
in Table \ref{tab:charges}.  Any $q$ or $\wt q$ dependence in the
coefficients $\{a,b,c,d\}$ must arise in (possibly inverse) powers of $q
\slash \wt q$.
As we expect the ring relations to be analytic in the size of the
$\pone$'s, such terms should not contribute.  
As a zeroth-order check of this statement, we can add $q$-
and $\wt q$-dependent terms to each coefficient $\{a,b,c,d\}$ -- in such a way
that the charges are respected -- and show that first-order additions necessarily
vanish.  
Consulting Table \ref{tab:charges}, we see that the combinations
\[
\epsilon_1\gamma_1,\quad
\epsilon_2\epsilon_3\gamma_2\gamma_3,\quad
\gamma_1^2 \epsilon_2\epsilon_3,\quad
\text{and}\quad
\epsilon_1^2 \gamma_2\gamma_3
\]
are of zero charge, so arbitrary functions of them may multiply any part of the
$q$, $\wt q$ terms.
Furthermore, we see that the only combinations of $q$ and $\wt q$
compatible with matching coefficient charges are powers of $q\slash \wt q$.
We perform our check by examining only $\sheaf O(\frac q{\wt q},\frac{\wt q}q)$
additions to the coefficients, with at most degree three polynomials in the
parameters multiplying $q$ and $\wt q$: for example, the allowed terms for $a$
would be 
\[
a\to 1+ \frac q {\wt q} \left (a_1 \gamma_1 \epsilon_1 + a_2 \gamma_2 \gamma_3 \right ) +
\frac {\wt q} q \left (a_3 \epsilon_2 \epsilon_3\right)
\]
for complex constants $a_i$.  One can check that each possible term must vanish
in order to satisfy the relations in \eqref{eq:relation2}. 

\begin{table}[t]
  \begin{tabular}{|cc|cc|cc|}
	 \hline
	 Quantity & Charge & Quantity & Charge& Quantity & Charge\\
	 \hline
	 $a$ & $0$                       & $\epsilon_1$           & $k - \wt k$ \Top & $q$ & $2k$\\
	 $b$ & $k - \wt k$               & $\gamma_1$             & $\wt k - k$ & $\wt q$ &$2\wt k$ \\
	 $c$ & $2(k - \wt k)\phantom{2}$ & $\epsilon_2\epsilon_3$ & $2(k - \wt k)\phantom 2$ &&  \\
	 $d$ & $2k$                      & $\gamma_2\gamma_3$     & $2(\wt k - k)\phantom 2$ &&  \\
	 \hline
  \end{tabular}
  \caption{Charge assignments for parameters and coefficients.}
  \label{tab:charges}
\end{table}

\section{Conclusions}

In this work, we computed correlators and deduced chiral ring relations in the
topological subsector of a $(0,2)$ supersymmetric gauged linear sigma model,
whose bosons in the geometric phase map to $\pones$ and whose left-moving
fermions couple to a deformation of the tangent bundle.  We find that the ring
relations in terms of deformation parameters $\epsilon_i, \gamma_i$ and a basis
$\{\psi, \wt \psi\}$ of topological operators are
\begin{alignat*}{4}
  \psi *\psi                          & +\epsilon _1 (\psi  *\wt \psi ) && {}-\epsilon _2 \epsilon _3 &&(\wt \psi *\wt \psi ) && =q\\
  \wt{\psi }*\wt{\psi } & +\gamma _1 (\psi *\wt \psi )    && {}-\gamma _2 \gamma _3 &&(\psi  *\psi)                           && =\wt{q}.
\end{alignat*}

Our correlation functions and ring relations match those computed in
\cite{Katz:2004nn} in the limit where the deformations match.  We were
neither able to verify nor refute the ring relations appearing in
\cite{Adams:2003zy}.  The difficulty arises in finding a change of basis
between our operators ($\psi$ and $\wt \psi$) and theirs ($X$ and $\wt X$).
Without loop corrections, it seems that in order to match results, the
basis change must involve square roots of polynomials in the deformation
parameters.\\

The ring relations derived in this work exhibit some curious features.  Note
that some deformation parameters appear quadratically: if we set all parameters
but $\epsilon_2$  and $\gamma_2$ to zero, the ring relations are the same as
those of the $(2,2)$ theory.  Indeed, the ring is apparently  insensitive to
those deformations parametrized by non-vanishing pairs
$\{\epsilon_i,\gamma_j\}$ among the parameters $\{ \epsilon_2, \epsilon_3,
\gamma_2,\gamma_3\}$, with the remaining two vanishing (for instance,
$\epsilon_2 = \gamma_2 = 0, \epsilon_3 \ne 0, \gamma_3 \ne 0$).  From the point
of view of \cite{Adams:2003zy}, the reason no modification to the ring
relations occurs is clear: for any given pair of non-vanishing parameters,
there is no combination such that the $U(1)$ charge of the term will not depend
on the $U(1)$ charges of the bosonic fields.  \\

The fact that each of these pairs of parameters give apparently non-trivial
theory deformations with the same chiral ring is quite interesting.  That there
are multiple pairs with this property shows that the invariance of the ring is
intrinsic to the problem, and not an artifact of the presentation of the
deformation.  \\

Another curiosity is the origin of the polynomial $\phi$ in the physics.  Since
it measures those deformations of the tangent bundle that are not bundles, it
is most likely that the GLSM of this deformation leads to a bad conformal field
theory.  Should the discrepancy between our results and those of the mirror
symmetry computation be satisfactorily explained, and the mirror map
understood, it would be interesting to compare the mirror theories in these
singular limits.  Presumably the mirror at these points becomes a badly behaved
CFT as well.

\section*{Acknowledgments}

We would like to thank
Allan Adams,
Anirban Basu,
Duncan Christie,
Ilarion Melnikov,
Sean Nowling,
and
Bernd Sturmfels
for
helpful discussions. 
We especially thank Eric Sharpe and Savdeep Sethi for their comments on
early drafts and many helpful discussions.  We also thank the organizers of
the 2006 Simons Workshop in Mathematics and Physics, where part of this work was
done.  This work was supported by NSF grants DMS 02-44412 and DMS 05-55678.

\appendix

\section{Lifts}
\label{sec:lifts}

On $\pones$, we compute the following elements of $C^0 \big(\sheaf O (-1,0)^2
\oplus \sheaf O (0,-1)^2 \big)$, which are mapped to $(1,0)$ via the matrix
\[\defMatrix.\] 

The lifts are, in terms of the quantity $\phi$ defined in \eqref{eq:phiDef},\\

\small
\begin{tabular}{rl}
  $U_{0,0}\Rightarrow$ &
\(\dfrac 1 {\phi\gamma_3 x_0 y_0^2}
\left(
\begin{array}{l}
 \gamma_2 y_1^2 (\gamma_2 \gamma_3 \epsilon_2 \epsilon_3 -1 ) - y_0 y_1 ( \gamma_1 + \gamma_2 \gamma_3 \epsilon_1 ) \\
 y_0 \epsilon_2 \gamma_3 \left [y_1 \gamma_2 (\gamma_2 \gamma_3 \epsilon_3 \epsilon_2 - 1) - y_0 (\gamma_1 - y_0 \gamma_2 \gamma_3
 \epsilon_1 )\right] \\
 x_0 \left[ y_0 \epsilon_2 \epsilon_3 \gamma_3 (\epsilon_1 \gamma_2 \gamma_3 + \gamma_1) + y_1 (1 -  \gamma_2 \epsilon_2 \epsilon_3
 \gamma_3) \right] \\
 x_1 \epsilon_2 \left[y_0 \left(\gamma_1+\gamma_2 \gamma_3 \epsilon_1\right)+y_1 \gamma_2 \left(1-\gamma_2 \gamma_3 \epsilon_2
 \epsilon_3\right)\right] 
 +x_0 y_1 \gamma_2 \epsilon_1 \left(1-\gamma_2 \gamma_3 \epsilon_2 \epsilon_3\right)\\
 \qquad+x_0 y_0 \left(\gamma_2 \gamma_3 \epsilon_1^2+\gamma_1 \epsilon_1-\left(\gamma_2 \gamma_3 \epsilon_2 \epsilon_3-1\right)^2\right)
 \end{array} \right) \) \\
  $U_{0,1}\Rightarrow$ &
  \(\dfrac 1 {\phi x_0 y_1^2}\left( \begin{array}{l}
 y_1^2 \left[\gamma_1 (\epsilon_2 \epsilon_3 \gamma_1+ \epsilon_1)
 + (\gamma_2 \gamma_3 \epsilon_2 \epsilon_3 -1) \right] -y_0 y_1  \gamma_3 (\epsilon_2 \epsilon_3 \gamma_1 + \epsilon_1)\\
 y_0  \epsilon_2 \gamma_3 \left [ - y_0 \gamma_3 ( \epsilon_1 + \epsilon_2 \epsilon_3 \gamma_1)
 + y_1 (\epsilon_2 \epsilon_3 \gamma_2 \gamma_3 -1 ) + y_1 \gamma_1(\epsilon_1+\gamma_1
 \epsilon_2\epsilon_3)\right]\\
 x_0 \gamma_3 \left [y_0 \gamma_3 \epsilon_2 \epsilon_3 \left( \gamma_1 \epsilon_2 \epsilon_3 + \epsilon_1 \right)
 + y_1 \epsilon_2 \epsilon_3 ( 1 - \gamma_2 \epsilon_2 \epsilon_3 \gamma_3 )
 + y_1 \epsilon_1 ( \epsilon_1 + \gamma_1 \epsilon_2 \epsilon_3 )\right]\\
 x_1 \epsilon_2 \left[y_0 \gamma_3 \left(\epsilon_1+\gamma_1 \epsilon_2 \epsilon_3\right)-y_1 \left(\gamma_1 \epsilon_1+\left(\gamma _1^2+\gamma_2
 \gamma_3\right) \epsilon_2 \epsilon_3-1\right)\right] \\
 \qquad-x_0 \left[y_0 \gamma_1 \gamma_3 \epsilon_2 \epsilon_3 \left(\epsilon_1+\gamma_1 \epsilon_2 \epsilon_3\right)+y_1 \epsilon_1
 \left(\gamma_1 \epsilon_1+\left(\gamma_1^2+\gamma_2 \gamma_3\right) \epsilon_2 \epsilon_3-1\right)\right]
\end{array} \right) \) \\
$U_{1,0}\Rightarrow$ & 
\(\dfrac 1 {\phi x_1 y_0}\left( \begin{array}{l}
 -y_1 \gamma_2 \epsilon_3 (1 - \gamma_2 \gamma_3 \epsilon_2 \epsilon_3)
 -y_0 \epsilon_3 ( \gamma_1 + \gamma_2 \gamma_3 \epsilon_1 )\\
 -y_1 \gamma_2 \epsilon_2 \epsilon_3 \left(\gamma_1+\gamma_2 \gamma_3 \epsilon_1\right) 
 +y_0 (\epsilon_2 \epsilon_3 \gamma_2\gamma_3  - 1 )
 +y_0 \epsilon_1 (\gamma_1 +  \epsilon_1 \gamma_2\gamma_3) \\
 x_1 \epsilon_2 \epsilon_3 \left(\gamma_1+\gamma_2 \gamma_3 \epsilon_1\right) +x_0 \epsilon_3 \left(1-\gamma_2 \gamma_3 \epsilon_2
 \epsilon_3\right)\\
 x_0 \gamma_2 \epsilon_3 \left(\epsilon_1+\gamma_1 \epsilon_2 \epsilon_3\right)
 +x_1 \gamma_2 \epsilon_2 \epsilon_3 \left(1-\gamma_2 \gamma_3 \epsilon_2 \epsilon_3\right)
\end{array} \right) \) \\
$U_{1,1}\Rightarrow$  &
\(\dfrac 1 {\phi x_1 y_1}\left(
\begin{array}{l}
  y_0 \gamma_3 \epsilon_3 \left(\gamma_2 \gamma_3 \epsilon_2 \epsilon_3-1\right)-
  y_1 \gamma_2 \gamma_3 \epsilon_3 \left(\epsilon_1+\gamma_1 \epsilon_2 \epsilon_3\right)\\
  y_1 \left[\gamma_1 \left(\epsilon_1+\gamma_1 \epsilon_2 \epsilon_3\right)+\gamma_2 \gamma_3 \left(\epsilon_1^2+\gamma_1 \epsilon_2 \epsilon_3 \epsilon_1+\epsilon_2 \epsilon_3\right)-1\right]\\
  \qquad - y_0 \epsilon_2 \epsilon_3 \gamma_3 \left(\gamma_1+\gamma_2 \gamma_3 \epsilon_1\right) \\
  x_0 \gamma_3 \epsilon_3 \left(\epsilon_1+\gamma_1 \epsilon_2 \epsilon_3\right)+x_1 \gamma_3 \epsilon_2 \epsilon_3 \left(1-\gamma_2 \gamma_3 \epsilon_2 \epsilon_3\right)\\
  x_1 \gamma_2 \gamma_3 \epsilon_2 \epsilon_3 \left(\epsilon_1+\gamma_1 \epsilon_2 \epsilon_3\right)
  +x_0 \epsilon_3 \left[1-\gamma_2 \gamma_3 \epsilon_2 \epsilon_3-\gamma_1 \left(\epsilon_1-\gamma_1 \epsilon_2 \epsilon_3\right)\right]\\
  \end{array}
\right) \)\\
\end{tabular}\\

\normalsize
Similarly, we compute lifts of $(0,1)$: \\
\small

\begin{tabular}{rl}
$U_{0,0}\Rightarrow$ & \(\dfrac 1 {\phi\gamma_3 x_0 y_0^2}\left( \begin{array}{l}
   y_1^2 \gamma_2 \left(\gamma_1+\gamma_2 \gamma_3 \epsilon_1\right)
  +y_0 y_1\left[\gamma_1 ( \gamma_1 +\gamma_2 \gamma_3 \epsilon_1 ) + \gamma_2 \gamma_3 \left(1-\gamma_2 \gamma_3 \epsilon_2 \epsilon_3\right)\right] \\
  y_0 \gamma_3 \epsilon_2 \left(y_1 \gamma_2 \left(\gamma_1+\gamma_2 \gamma_3 \epsilon_1\right)
  +y_0 \left[\gamma_1 ( \gamma_1 +\gamma_2 \gamma_3 \epsilon_1 ) + \gamma_2 \gamma_3 \left(1-\gamma_2 \gamma_3 \epsilon_2 \epsilon_3\right)\right]
  \right) \\
  x_0 y_0 \gamma_3 \left[\epsilon_1 \left(\gamma_1+\gamma_2 \gamma_3 \epsilon_1\right)+\gamma_2 \gamma_3 \epsilon_2 \epsilon_3-1\right]-x_0 y_1
  \left(\gamma_1+\gamma_2 \gamma_3 \epsilon_1\right)\\
-x_0 y_1 \gamma_2 \epsilon_1 \left(\gamma_1+\gamma_2 \gamma_3 \epsilon_1\right)-x_1 y_1 \gamma_2 \epsilon_2 \left(\gamma_1+\gamma_2 \gamma_3 \epsilon_1\right)\\
\qquad-x_0 y_0 \gamma_1 \left[\epsilon_1 \left(\gamma_1+\gamma_2 \gamma_3 \epsilon_1\right)+\gamma_2 \gamma_3 \epsilon_2 \epsilon_3-1\right]\\
\qquad-x_1 y_0 \epsilon_2 \left[\gamma_1 \left(\gamma_1+\gamma_2 \gamma_3 \epsilon_1\right)+\gamma_2 \gamma_3 \left(1-\gamma_2 \gamma_3 \epsilon_2 \epsilon_3\right)\right]
  \end{array} \right) \) \\
  $U_{0,1}\Rightarrow$ &
  \(\dfrac 1 {\phi x_0 y_1^2}\left(
  \begin{array}{l}
         \gamma_3 y_1 \left[ y_1 \gamma_2 \left(\epsilon_1+\gamma_1 \epsilon_2 \epsilon_3\right)
         +y_0 \left(1-\gamma_2 \gamma_3 \epsilon_2 \epsilon_3\right) \right]\\
         y_0 \gamma_3^2 \epsilon_2 \left[y_1 \gamma_2 \left(\epsilon_1+\gamma_1 \epsilon_2 \epsilon_3\right)
         +y_0 \left(1-\gamma_2 \gamma_3 \epsilon_2 \epsilon_3\right)\right]\\
         x_0 \gamma_3 \left[y_0 \gamma_3 \epsilon_2 \epsilon_3 \left(\gamma_2 \gamma_3 \epsilon_2 \epsilon_3-1\right)-y_1 \left(\epsilon_1+\gamma_1 \epsilon_2 \epsilon_3\right)\right]\\
         y_0 \gamma_3 \epsilon_2 \left(x_1-x_0 \gamma_1 \epsilon_3\right) \left(\gamma_2 \gamma_3 \epsilon_2 \epsilon_3-1\right)
         -y_1 x_1 \gamma_2 \gamma_3 \epsilon_2 \left(\epsilon_1+\gamma_1 \epsilon_2 \epsilon_3\right)\\
         \qquad+y_1 x_0 \left[\epsilon_2 \epsilon_3 \gamma_1^2+\epsilon_1 \gamma_1-\left(\gamma_2 \gamma_3 \epsilon_2 \epsilon_3-1\right)^2\right]
  \end{array} \right)\) \\
  $U_{1,0}\Rightarrow$ &
  \(\dfrac 1 {\phi x_1 y_0}\left(
  \begin{array}{l}
         y_1 \epsilon_3\gamma_2 \left(\gamma_1+\gamma_2 \gamma_3 \epsilon_1\right) 
         +y_0 \epsilon_3\left[\gamma_1 ( \gamma_1 +\gamma_2 \gamma_3 \epsilon_1 ) + \gamma_2 \gamma_3 \left(1-\gamma_2 \gamma_3 \epsilon_2 \epsilon_3\right)\right] \\
         y_0 \left[\gamma_1 \left(1-\gamma_1 \epsilon_1\right)+\gamma_2 \gamma_3 \epsilon_1 \left(\gamma_2 \gamma_3 \epsilon_2 \epsilon_3-\gamma_1 \epsilon_1\right)\right]\\
         \qquad-y_1 \gamma_2 \left[\epsilon_1 \left(\gamma_1+\gamma_2 \gamma_3 \epsilon_1\right)+\gamma_2 \gamma_3 \epsilon_2 \epsilon_3-1\right]\\
         x_1 \left[\epsilon_1 \left(\gamma_1+\gamma_2 \gamma_3 \epsilon_1\right)+\gamma_2 \gamma_3 \epsilon_2 \epsilon_3-1\right]
         -x_0 \left(\gamma_1+\gamma_2 \gamma_3 \epsilon_1\right) \epsilon_3\\
         x_0 \gamma_2 \epsilon_3 \left(\gamma_2 \gamma_3 \epsilon_2 \epsilon_3-1\right)-x_1 \gamma_2\epsilon_2 \epsilon_3 \left(\gamma_1+\gamma_2 \gamma_3 \epsilon_1\right) 
  \end{array}
  \right) \) \\
  $U_{1,1}\Rightarrow$ &
\(\dfrac 1 {\phi x_1 y_1}\left(
\begin{array}{l}
         y_0 \epsilon_3\gamma_3 \left(\gamma_1+\gamma_2 \gamma_3 \epsilon_1\right) +y_1 \epsilon_3\gamma_2 \gamma_3 \left(1-\gamma_2 \gamma_3 \epsilon_2 \epsilon_3\right) \\
         y_1 \gamma_2 \gamma_3 \left(\gamma_1+\gamma_2 \gamma_3 \epsilon_1\right) \epsilon_2 \epsilon_3
         +y_0 \gamma_3 \left[1-\gamma_2 \gamma_3 \epsilon_2 \epsilon_3-\epsilon_1 \left(\gamma_1+\gamma_2 \gamma_3 \epsilon_1\right)\right]\\
         x_0 \gamma_3 \epsilon_3 \left(\gamma_2 \gamma_3 \epsilon_2 \epsilon_3-1\right)-x_1 \epsilon_2 \epsilon_3\gamma_3 \left(\gamma_1+\gamma_2 \gamma_3 \epsilon_1\right)  \\
         x_1 \left[\gamma_2 \gamma_3 \epsilon_2 \epsilon_3-1+
         (\gamma_1 +\gamma_2 \gamma_3 \epsilon_1 )(\epsilon_1+\gamma_1 \epsilon_2 \epsilon_3)\right]
         -x_0 \gamma_2 \gamma_3 \epsilon_3 \left(\epsilon_1+\gamma_1 \epsilon_2 \epsilon_3\right)
\end{array} \right) \)\\
\end{tabular}

\normalsize

\section{Correlators}
\label{sec:correlators}

Effectuation of the algorithm detailed in \S \ref{sec:algorithm} yields the
following correlators; 
a
Degree (0,0):
\begin{equation}
  \begin{split}
         \langle \psi \psi \rangle &= 
         \frac 1 \phi ( \epsilon_1+\gamma_1 \epsilon_2 \epsilon_3)\\
         \langle \psi \wt \psi \rangle &= 
         \frac 1 \phi ( \gamma_2 \gamma_3 \epsilon_2 \epsilon_3-1)\\
         \langle \wt \psi \wt \psi \rangle &=  
         \frac 1 \phi (\gamma_1 + \epsilon_1\gamma_2 \gamma_3 )
  \end{split}
  \label{eq:degree00correlators}
\end{equation}

Degree (1,0):
\begin{equation}
  \begin{split}
         \langle \psi \psi \psi \psi \rangle_{1,0} &=  \frac {1} {\phi^2}
         \left(\epsilon _1+\gamma _1 \epsilon _2 \epsilon _3\right) \left[\gamma_1
         (\epsilon_1+ \gamma_1\epsilon_2 \epsilon_3) + 2 (\gamma_2 \gamma_3
         \epsilon_2 \epsilon_3-1)\right]\\
         \langle \psi \psi \psi \wt \psi \rangle_{1,0} &=  \frac {1} {\phi^2}
         \left[ (\gamma_2 \gamma_3 \epsilon_2 \epsilon_3 - 1)^2 + \gamma_2 \gamma_3
         \left(\epsilon_1+\gamma_1 \epsilon_2 \epsilon_3\right)^2\right]\\
         \langle \psi \psi \wt \psi \wt \psi \rangle_{1,0} &=  \frac {1} {\phi^2}
         \left(\gamma_2 \gamma_3 \epsilon_2 \epsilon_3-1\right)
         \left[2 \left(\gamma_1+\gamma_2 \gamma_3 \epsilon_1\right)-\gamma_1
         \left(1-\gamma_2 \gamma_3 \epsilon_2 \epsilon_3\right)\right]\\
         \langle \psi \wt \psi \wt \psi \wt \psi \rangle_{1,0} &= 
         \frac {1} {\phi^2} \left[ \left(\gamma_1+\gamma_2 \gamma_3
         \epsilon_1\right)^2+\gamma_2 \gamma_3 \left(\gamma_2 \gamma_3
         \epsilon_2 \epsilon_3-1\right)^2 \right] \\
         \langle \wt \psi \wt \psi \wt \psi \wt \psi \rangle_{1,0} &=  \frac {-1} {\phi^2}
         \left ( \gamma_1 + \epsilon_1 \gamma_2 \gamma_3 \right) 
			\left[ \gamma_1 \left(\gamma_1+\gamma_2 \gamma_3 \epsilon_1\right)-
         2 \gamma_2 \gamma_3 \left(\gamma_2 \gamma_3 \epsilon_2
         \epsilon_3-1\right) \right]
  \end{split}
  \label{eq:degree10correlators}
\end{equation}

Degree (0,1):
\begin{equation}
  \begin{split}
         \langle \psi \psi \psi \psi \rangle_{0,1} &=  \frac {-1} {\phi^2}
         \left( \epsilon_1 + \gamma_1 \epsilon_2 \epsilon_3 \right)
			\left[ \epsilon_1 \left(\epsilon_1+\gamma_1 \epsilon_2 \epsilon_3\right)
         +2 \epsilon_2 \epsilon_3 \left(1-\gamma_2 \gamma_3 \epsilon_2
         \epsilon_3 \right) \right] \\
         \langle \psi \psi \psi \wt \psi \rangle_{0,1} &=  \frac {1} {\phi^2} 
         \left [\left(\epsilon_1+\gamma_1 \epsilon_2
         \epsilon_3\right)^2+\epsilon_2 \epsilon_3 \left(\gamma_2 \gamma_3
         \epsilon_2 \epsilon_3-1\right)^2\right]\\
         \langle \psi \psi \wt \psi \wt \psi \rangle_{0,1} &=  \frac {1} {\phi^2}
         \left(\gamma_2 \gamma_3 \epsilon_2 \epsilon_3-1\right)
         \left[ \epsilon_1 (\gamma_2 \gamma_3 \epsilon_2 \epsilon_3 - 1) + 2 (
         \epsilon_1+ \gamma_1 \epsilon_2 \epsilon_3)\right]\\
         \langle \psi \wt \psi \wt \psi \wt \psi \rangle_{0,1} &=  \frac {1} {\phi^2}
         \left[\epsilon_2 \epsilon_3 \left(\gamma_1+\gamma_2 \gamma_3 \epsilon_1\right)^2+\left(1-\gamma_2 \gamma_3 \epsilon_2 \epsilon_3\right)^2
         \right]\\
         \langle \wt \psi \wt \psi \wt \psi \wt \psi \rangle_{0,1} &=  \frac {1} {\phi^2}
			\left (\gamma_1 + \epsilon_1 \gamma_2 \gamma_3 \right)
			\left[ \epsilon_1 \left(\gamma_1+\gamma_2 \gamma_3 \epsilon_1\right)+2
         \left(\gamma_2 \gamma_3 \epsilon_2 \epsilon_3-1\right) \right] \\
  \end{split}
  \label{eq:degree01correlators}
\end{equation}

The overall four-point correlation functions are then of the form 
\begin{equation}
  \langle \psi\psi\psi\psi \rangle = 
  q     \cdot \langle \psi\psi\psi\psi \rangle_{1,0} +
  \wt q \cdot \langle \psi\psi\psi\psi \rangle_{0,1} .
  \label{eq:fourPointFunction}
\end{equation}

Six-point functions from the $(0,2)$, $(2,0)$, and $(1,1)$ sectors involve too
many terms to express here.  A Mathematica notebook containing all computed
correlators may be obtained from the authors.

\bibliography{deformation}

\end{document}